\documentclass[structabstract]{aa}

\usepackage{graphicx}
\usepackage[usenames]{color} 

\begin{document}

\title{
B fields in OB stars (BOB): \\
Concluding the FORS\,2 observing campaign\thanks{
Based on observations made with ESO Telescopes at the La Silla Paranal Observatory
under programme ID~191.D-0255(E,G).}
}

\author{
M.~Sch\"oller\inst{1}
	\and
S.~Hubrig\inst{2}
	\and
L.~Fossati\inst{3,4}
	\and
T.~A.~Carroll\inst{2}
	\and
M.~Briquet\inst{5,2}
	\and
L.~M.~Oskinova\inst{6}
	\and
S.~J\"arvinen\inst{2}
	\and
I.~Ilyin\inst{2}
	\and
N.~Castro\inst{7,4}
	\and
T.~Morel\inst{5}
	\and
N.~Langer\inst{4}
	\and
N.~Przybilla\inst{8}
	\and
M.-F.~Nieva\inst{8}
	\and
A.~F.~Kholtygin\inst{9}
	\and
H.~Sana\inst{10}
	\and
A.~Herrero\inst{11,12}
	\and
R.~H.~Barb\'a\inst{13}
	\and
A.~de~Koter\inst{14}
	\and
the~BOB~collaboration
}

   \institute{
European Southern Observatory,
Karl-Schwarzschild-Str.~2,
85748 Garching, Germany\\
\email{mschoell@eso.org}
	\and
Leibniz-Institut f\"ur Astrophysik Potsdam (AIP),
An der Sternwarte~16,
14482~Potsdam, Germany
	\and
\"Osterreichische Akademie der Wissenschaften, Institut f\"ur Weltraumforschung, Schmiedlstra\ss{}e~6, 8042~Graz, Austria
	\and
Argelander-Institut f\"ur Astronomie der Universit\"at Bonn, Auf dem H\"ugel 71, 53121~Bonn, Germany
	\and
Space sciences, Technologies and Astrophysics Research (STAR) Institute, Universit\'e de Li\`ege, Quartier Agora, All\'ee du 6 Ao\^ut 19c, B\^at.~B5C, 4000~Li\`ege, Belgium
	\and
Universit\"at Potsdam, Institut f\"ur Physik und Astronomie, Karl-Liebknecht-Str.~24/25, 14476 Potsdam, Germany
	\and
Department of Astronomy, University of Michigan, 1085 S.~University Avenue, Ann~Arbor, MI~48109-1107, USA
	\and
Institut f\"ur Astro- und Teilchenphysik, Universit\"at Innsbruck, Technikerstr.~25/8, 6020~Innsbruck, Austria
	\and
St.~Petersburg State University, Universitetski pr.~28, St.~Petersburg 198504, Russia
	\and
Institute of Astrophysics, KU Leuven, Celestijnlaan 200D, 3001~Leuven, Belgium
	\and
Instituto de Astrof\'isica de Canarias, C/ V\'ia L\'actea s/n, 38205 La Laguna, Spain
	\and
Universidad de La Laguna, Departamento de Astrof\'isica, Avda.~Francisco Sanchez 2, 38206 La Laguna, Spain
	\and
Departamento de F\'isica y Astronom\'ia, Universidad de La Serena, Av.~Juan Cisternas 1200 Norte, La Serena, Chile
	\and
Anton Pannekoek Institute for Astronomy, University of Amsterdam, Science Park 904, PO Box 94249, 1090 GE, Amsterdam, The Netherlands
             }

   \date{Received September 15, 1996; accepted March 16, 1997}

 
  \abstract
   {}
   {
The ``B fields in OB stars'' (BOB) collaboration is based on an
ESO Large Programme,
to study the occurrence rate, properties, and ultimately the origin of magnetic fields in massive stars.
   }
   {
In the framework of this programme,
we carried out low-resolution spectropolarimetric observations
of a large sample of massive stars using FORS\,2 installed at the ESO VLT 8-m telescope.
   }
   {
We determined the magnetic field values with two completely independent reduction and analysis pipelines.
Our in-depth study of the magnetic field measurements
shows that differences between our two pipelines are usually well within 3$\sigma$ errors.
From the 32 observations of 28 OB stars,
we were able to monitor the magnetic fields in CPD\,$-$57$^{\circ}$\,3509 and HD\,164492C,
confirm the magnetic field in HD\,54879,
and detect a magnetic field in CPD\,$-$62$^{\circ}$\,2124.
We obtain a magnetic field detection rate of $6\pm3\%$ for the full sample of 69 OB stars observed
with FORS\,2 within the BOB programme.
For the pre-selected objects with a $v\,\sin\,i$ below 60\,km\,s$^{-1}$, we obtain a magnetic field
detection rate of $5\pm5\%$.
We also discuss X-ray properties and multiplicity of the objects in our FORS\,2 sample
with respect to the magnetic field detections.
   }
   {}


   \maketitle

\section{Introduction}
\label{sect:intro}

Magnetic fields may have a key influence on the evolution of massive stars.
Magnetized winds can spin down the star by applying a torque at the surface 
(ud-Doula et al.\ \cite{udDoula2008}; Meynet et al.\ \cite{Meynet2011})
and they can mediate angular momentum transport in the stellar interior 
(Heger et al.\ \cite{Heger2005}; Maeder \& Meynet \cite{MaederMeynet2005}).
Understanding these processes will provide strong constraints on the
role of rotationally induced mixing in massive stars (Brott et al.\ \cite{Brott2011}),
and on the frequency of long gamma-ray bursts (Yoon et al.\ \cite{Yoon2006})
and massive black hole mergers (Mandel \& de Mink \cite{MandeldeMink2016};
Marchant et al.\ \cite{Marchant2016}).
Dedicated surveys targeting the detection and the characterization of magnetic fields
in massive stars have started only in recent years
(Wade et al.\ \cite{Wade2016}; Morel et al.\ \cite{Morel2014,Morel2015}),
leading to an increase in the number of massive stars with characterized magnetic fields.

The ``B  fields  in OB stars'' (BOB) collaboration, established
in 2013, aims at investigating the frequency and strength distribution of magnetic fields
in OB stars using spectropolarimetric observations, concentrating mainly on slow rotators.
Data are obtained in the framework of the ESO
Large Programme 191.D-0255, scheduled on the 8-m Very Large Telescope (VLT) on Paranal and the
ESO 3.6-m~telescope on La~Silla.
The instruments used for the spectropolarimetric observations are the
FOcal Reducer low dispersion Spectrograph
(FORS\,2; Appenzeller et al.\ \cite{Appenzeller1998}) and
the High Accuracy Radial velocity Planet Searcher polarimeter (HARPSpol; Snik et al.\ \cite{Snik2008}).
Details about the aims and the implications of the first discoveries by the BOB survey
are presented by Morel et al.\ (\cite{Morel2014,Morel2015}) and Fossati et al.\ (\cite{Fossati2015}).

In this work, we present the most recent FORS\,2 observations of 28~stars,
concluding the low-resolution spectropolarimetric campaign within the BOB survey.
We describe the observations and data reduction in Sect.~\ref{sect:observations},
present the magnetic field measurements in Sect.~\ref{sect:results},
give an extensive overview about the statistics and quality of the measurements in Sect.~\ref{sect:analysis},
discuss some object properties in Sect.~\ref{sect:indicators},
and finally conclude with a discussion in Sect.~\ref{sect:discussion}.

\section{Observations and data reduction}
\label{sect:observations}

The BOB collaboration is mainly targeting OB dwarfs and giants with low projected
rotation velocities.
A full description of the target selection criteria was presented by Fossati et al.\ (\cite{Fossati2015}),
who published FORS\,2 magnetic field measurements of 50 O- and early B-type stars
obtained in runs A and C.
Since the presence of weak magnetic fields in Be-type stars is not yet established,
we additionally carried out observations of four Be stars,
three of which were selected on the basis of their hard X-ray spectra (see Sect.~\ref{sect:xrays}).

\begin{table}
\centering
\caption{
Objects studied with FORS\,2 during the runs in 2014 June
and 2015 March.
Objects marked by an asterisk were also observed in runs A and C
(Fossati et al.\ \cite{Fossati2015}).
}
\label{tab:objects}
\begin{tabular}{lcr@{.}l}
\hline
\hline
\multicolumn{1}{c}{Object} &
\multicolumn{1}{c}{Spectral Type} &
\multicolumn{2}{c}{$m_{\rm V}$} \\
\hline
BD\,$-$12$^{\circ}$\,4982         & B0\,II               &  9&22\\
CD\,$-$22$^{\circ}$\,12513        & B0\,V                & 10&01\\
CPD\,$-$57$^{\circ}$\,3509$^\ast$ & B2\,IV               & 10&68\\
CPD\,$-$62$^{\circ}$\,2124        & B2\,IV               & 10&99\\
HD\,54879$^\ast$        & O9.7\,V              &  7&64\\
HD\,56779               & B2\,IV-V             &  5&02\\
HD\,72754$^\ast$        & B2Ia.pshe            &  6&85\\
HD\,75759               & O9\,V\,+\,B0\,V      &  5&99\\
HD\,95568$^\ast$        & O9/B0\,V             &  9&60\\
HD\,97991$^\ast$        & B2\,II               &  7&40\\
HD\,110432              & B2\,pe               &  5&31\\
HD\,118198$^\ast$       & O9.7\,III            &  8&47\\
HD\,120324              & B2\,Vnpe             &  3&35\\
HD\,120991              & B2\,IIne             &  6&08\\
HD\,156134              & B0\,Ib               &  8&06\\
HD\,156233              & B0                   &  9&10\\
HD\,156234              & B0\,III              &  7&74\\
HD\,156292              & O9.7\,III            &  7&51\\
HD\,164492B             & B2\,Vnn              & 10&52\\
HD\,164492C$^\ast$      & B1\,V                &  8&66\\
HD\,164536              & O7.5\,V(n)z          &  7&11\\
HD\,164704              & B2\,II               &  8&16\\
HD\,164816              & O9.5\,V\,+\,B0\,V    &  7&09\\
HD\,164844              & B1/B2\,III           &  8&29\\
HD\,165052              & O7\,Vz\,+\,O7.5\,Vz  &  6&87\\
HD\,166033              & B1\,V                &  9&60\\
HD\,315032              & B2\,Vne              &  9&19\\
HD\,345439              & B1/B2\,V             & 11&26\\
\hline
\end{tabular}
\end{table}

We conducted 32 new spectropolarimetric observations of 28~OB stars
in visitor mode with FORS\,2 during 2.5~nights each on
2014 June 1--3 and 2015 March 16--18, as part
of runs E and G of our observing campaign.
No data were obtained on 2015 March 16 and 18, due to high humidity and clouds, respectively.
All other nights had severe issues with clouds and/or strong winds.
The observed objects are listed in Table~\ref{tab:objects}, together with their spectral type and
their V~magnitude.
Main sources of the spectral classification are
the Galactic O-Star Spectroscopic Survey (Sota et al.\ \cite{Sota2014}),
the Michigan Catalogue of HD stars (see Houk \cite{Houk1994}), and previous BOB publications.
The primary source for the photometry is the catalogue of Mermilliod (\cite{Mermilliod2006}).
For seven of these objects, observations were already carried out in earlier runs with FORS\,2 and 
were reported by Fossati et al.\ (\cite{Fossati2015}).
We reobserved these sources either to check earlier marginal magnetic field detections
or to obtain magnetic field measurements at different phases of the object's -- usually unknown -- rotation cycle. 
In total, 71 different objects were observed with FORS\,2 within the BOB programme, in 134 observations.

FORS\,2 is a multi-mode instrument equipped with polarization-analyzing optics
comprising super-achromatic half-wave and quarter-wave phase retarder plates,
and a Wollaston prism with a beam divergence of 22$\arcsec$ in standard-resolution mode.
We used the GRISM 600B and the narrowest available slit width
of 0$\farcs$4 to obtain a spectral resolving power of $R\sim2000$.
The observed spectral range from 3250 to 6215\,\AA{} includes all Balmer lines
apart from H$\alpha$, and numerous \ion{He}{i} lines.
The position angle of the retarder waveplate was alternated in the sequence
$-45^{\circ}$$+45^{\circ}$$+45^{\circ}$$-45^{\circ}$.
Between two to six such sequences of exposures were combined to form a single polarimetric observation.
For the observations, we used a non-standard readout mode with low
gain (200kHz,1$\times$1,low), which provides a broader dynamic range, hence
allowing us to reach a higher signal-to-noise ratio (S/N) in the individual exposures,
which is especially advantageous for bright targets.

The determination of the mean longitudinal magnetic field using low-resolution FORS\,1/2 spectropolarimetry
has been described in detail by two different groups 
(Bagnulo et al.\ \cite{Bagnulo2002,Bagnulo2009,Bagnulo2012};
Hubrig et al.\ \cite{Hubrig2004a,Hubrig2004b,Hubrig2016};
Sch\"oller et al., in preparation).
The $V/I$ spectrum is calculated using:

\begin{equation}
\frac{V}{I} = \frac{1}{2} \left\{
\left( \frac{f^{\rm o} - f^{\rm e}}{f^{\rm o} + f^{\rm e}} \right)_{-45^{\circ}} -
\left( \frac{f^{\rm o} - f^{\rm e}}{f^{\rm o} + f^{\rm e}} \right)_{+45^{\circ}} \right\}
\end{equation}

\noindent
where $+45^{\circ}$ and $-45^{\circ}$ indicate the position angle of the
quarter-wave plate and $f^{\rm o}$ and $f^{\rm e}$ are the fluxes of the ordinary and
extraordinary beams, respectively.

For low-resolution FORS\,2 spectra, the mean longitudinal magnetic field $\left<B_{\rm z}\right>$
is usually diagnosed from the slope of a linear regression of $V/I$ versus the quantity
$-g_{\rm eff} \Delta\lambda_z \lambda^2 \frac{1}{I} \frac{{\mathrm d}I}{{\mathrm d}\lambda} \left<B_{\rm z}\right > + V_0/I_0$,
where $V$ is the Stokes parameter that measures the circular polarization,
$I$ is the intensity observed in unpolarized light,
$g_{\rm eff}$ is the effective Land\'e factor,
$\lambda$ is the wavelength,
${{\rm d}I/{\rm d}\lambda}$ is the derivative of Stokes $I$,
and $V_0/I_0$ denotes the instrumental polarization.
The diagnostic $\left<N_{\rm z}\right>$ parameter was calculated following the formalism
of Bagnulo et al.\ (\cite{Bagnulo2009}).

To identify systematic differences that might exist when the FORS\,2 data are treated by different groups,
the mean longitudinal magnetic field, $\left< B_{\rm z}\right>$, was derived for all stars by two groups separately, using
independent reduction packages.
Details of and differences between the two reduction and analysis packages are described in detail
by Fossati et al.\ (\cite{Fossati2015}).

\section{Results}
\label{sect:results}

\begin{table*}
\centering
\caption{
Results from our spectropolarimetric observations with FORS\,2 in 2014 June and 2015 March.
Names of objects with magnetic field detections are highlighted in bold face.
}
\label{tab:MFs}
\begin{tabular}{llrl| r @{$\pm$} l @{/} r r @{$\pm$} l @{/} r|r @{$\pm$} l @{/} r r @{$\pm$} l @{/} r}
\hline
\hline
\multicolumn{1}{c}{Object} &
\multicolumn{1}{c}{MJD} &
\multicolumn{1}{c}{S/N} &
\multicolumn{1}{c}{Group} &
\multicolumn{6}{|c}{H lines} &
\multicolumn{6}{|c}{Whole spectrum} \\
\multicolumn{1}{c}{} &
\multicolumn{1}{c}{$n_{\rm exp}$/$t_{\rm exp}$ [s]} &
\multicolumn{1}{c}{} &
\multicolumn{1}{c}{} &
\multicolumn{2}{|c}{$\left<B_{\rm z}\right>$ [G]} &
\multicolumn{1}{c}{$\sigma$} &
\multicolumn{2}{c}{$\left<N_{\rm z}\right>$ [G]} &
\multicolumn{1}{c}{$\sigma$} &
\multicolumn{2}{|c}{$\left<B_{\rm z}\right>$ [G]} &
\multicolumn{1}{c}{$\sigma$} &
\multicolumn{2}{c}{$\left<N_{\rm z}\right>$ [G]} &
\multicolumn{1}{c}{$\sigma$} \\
\hline
HD\,72754 & 56809.94942 &  	 3416 & Potsdam & $-$50 & 36 & 1.4 & $-$29 & 33 & 0.9 & $-$29 & 27 & 1.1 & $-$16 & 24 & 0.7 \\ 	 & \multicolumn{1}{c}{24/320}	& 3308 & Bonn & $-$53 & 26 & 2.0 & $-$36 & 23 & 1.6 & $-$17 & 16 & 1.1 & $-$41 & 15 & 2.7 \\ 
HD\,72754 & 56810.94153 &  	 2906 & Potsdam & $-$65 & 47 & 1.4 & 17 & 28 & 0.6 & $-$55 & 30 & 1.8 & 31 & 20 & 1.6 \\ 	 & \multicolumn{1}{c}{12/203}	& 2779 & Bonn & $-$38 & 28 & 1.4 & $-$24 & 25 & 1.0 & $-$21 & 18 & 1.2 & 14 & 17 & 0.8 \\ 
CPD\,$-$57$^{\circ}$\,3509 & 56809.97981 &  	 2025 & Potsdam & $-$19 & 71 & 0.3 & $-$28 & 86 & 0.3 & 88 & 54 & 1.6 & $-$45 & 59 & 0.8 \\ 	 & \multicolumn{1}{c}{8/4400}	& 1943 & Bonn & $-$71 & 71 & 1.0 & $-$58 & 77 & 0.8 & 40 & 46 & 0.9 & $-$51 & 47 & 1.1 \\ 
{\bf CPD\,$-$57$^{\circ}$\,3509} & 56810.96178 &  	 2348 & Potsdam & 979 & 68 & 14.4 & $-$108 & 77 & 1.4 & 920 & 48 & 19.2 & 2 & 50 & 0.0 \\ 	 & \multicolumn{1}{c}{8/4400}	& 2289 & Bonn & 1049 & 69 & 15.2 & $-$90 & 61 & 1.5 & 943 & 41 & 23.0 & 1 & 39 & 0.0 \\ 
{\bf CPD\,$-$57$^{\circ}$\,3509} & 57099.06734 &  	 1826 & Potsdam & 582 & 99 & 5.9 & $-$75 & 101 & 0.7 & 671 & 62 & 10.8 & $-$33 & 61 & 0.5 \\ 	 & \multicolumn{1}{c}{8/4500}	& 1791 & Bonn & 607 & 98 & 6.2 & $-$4 & 89 & 0.0 & 734 & 56 & 13.1 & 8 & 55 & 0.1 \\ 
HD\,118198 & 56810.04647 &  	 2628 & Potsdam & $-$36 & 60 & 0.6 & 46 & 68 & 0.7 & $-$84 & 34 & 2.5 & $-$14 & 38 & 0.4 \\ 	 & \multicolumn{1}{c}{8/800}	& 2583 & Bonn & $-$65 & 58 & 1.1 & 13 & 57 & 0.2 & $-$66 & 37 & 1.8 & $-$43 & 37 & 1.2 \\ 
{\bf HD\,164492C} & 56810.06846 &  	 2829 & Potsdam & 651 & 85 & 7.7 & $-$52 & 73 & 0.7 & 540 & 72 & 7.5 & $-$59 & 49 & 1.2 \\ 	 & \multicolumn{1}{c}{8/1520}	& 2775 & Bonn & 756 & 74 & 10.2 & $-$3 & 57 & 0.1 & 601 & 49 & 12.3 & $-$32 & 42 & 0.8 \\ 
{\bf HD\,164492C} & 57099.31332 &  	 2666 & Potsdam & $-$495 & 92 & 5.4 & $-$95 & 89 & 1.1 & $-$544 & 66 & 8.2 & $-$31 & 62 & 0.5 \\ 	 & \multicolumn{1}{c}{8/1320}	& 2612 & Bonn & $-$574 & 88 & 6.5 & $-$81 & 78 & 1.0 & $-$551 & 56 & 9.8 & $-$35 & 51 & 0.7 \\ 
HD\,164492B & 56810.09776 &  	 1530 & Potsdam & $-$27 & 150 & 0.2 & 273 & 177 & 1.5 & $-$44 & 130 & 0.3 & $-$199 & 110 & 1.8 \\ 	 & \multicolumn{1}{c}{8/4400}	& 1547 & Bonn & $-$59 & 138 & 0.4 & 4 & 133 & 0.0 & 26 & 115 & 0.2 & $-$2 & 119 & 0.0 \\ 
HD\,164704 & 56810.16040 &  	 2667 & Potsdam & 23 & 83 & 0.3 & $-$46 & 74 & 0.6 & 18 & 46 & 0.4 & 3 & 44 & 0.1 \\ 	 & \multicolumn{1}{c}{8/560}	& 2596 & Bonn & $-$43 & 61 & 0.7 & $-$97 & 55 & 1.8 & 32 & 38 & 0.8 & $-$53 & 36 & 1.5 \\ 
HD\,166033 & 56810.17880 &  	 2702 & Potsdam & $-$90 & 94 & 1.0 & $-$63 & 88 & 0.7 & $-$131 & 58 & 2.3 & $-$103 & 53 & 1.9 \\ 	 & \multicolumn{1}{c}{8/800}	& 2655 & Bonn & 32 & 65 & 0.5 & 14 & 64 & 0.2 & 5 & 44 & 0.1 & 2 & 43 & 0.0 \\ 
BD\,$-$12\,4982 & 56810.19963 &  	 2510 & Potsdam & 130 & 130 & 1.0 & 51 & 111 & 0.5 & 271 & 52 & 5.2 & $-$40 & 53 & 0.8 \\ 	 & \multicolumn{1}{c}{8/2700}	& 2113 & Bonn & 97 & 98 & 1.0 & 148 & 84 & 1.8 & 266 & 57 & 4.7 & $-$21 & 47 & 0.4 \\ 
HD\,345439 & 56810.24718 &  	 1140 & Potsdam & 57 & 153 & 0.4 & 58 & 136 & 0.4 & $-$22 & 116 & 0.2 & 64 & 103 & 0.6 \\ 	 & \multicolumn{1}{c}{8/4800}	& 1678 & Bonn & 9 & 132 & 0.1 & 25 & 108 & 0.2 & $-$4 & 79 & 0.1 & 61 & 74 & 0.8 \\ 
CD\,$-$22\,12513 & 56810.31559 &  	 2706 & Potsdam & $-$105 & 53 & 2.0 & $-$53 & 63 & 0.8 & $-$106 & 35 & 3.0 & $-$52 & 41 & 1.3 \\ 	 & \multicolumn{1}{c}{8/2400}	& 2585 & Bonn & $-$89 & 52 & 1.7 & $-$82 & 46 & 1.8 & $-$77 & 36 & 2.1 & $-$79 & 35 & 2.3 \\ 
HD\,164536 & 56810.35596 &  	 3472 & Potsdam & 108 & 77 & 1.4 & $-$64 & 85 & 0.8 & 56 & 63 & 0.9 & $-$48 & 72 & 0.7 \\ 	 & \multicolumn{1}{c}{12/317}	& 3331 & Bonn & 193 & 68 & 2.8 & $-$93 & 68 & 1.4 & 71 & 53 & 1.3 & $-$80 & 53 & 1.5 \\ 
HD\,164816 & 56810.37405 &  	 3576 & Potsdam & $-$60 & 68 & 0.9 & $-$34 & 62 & 0.5 & $-$60 & 45 & 1.3 & $-$98 & 44 & 2.2 \\ 	 & \multicolumn{1}{c}{12/480}	& 3469 & Bonn & $-$35 & 51 & 0.7 & $-$22 & 46 & 0.5 & $-$40 & 35 & 1.1 & $-$76 & 33 & 2.3 \\ 
HD\,165052 & 56810.39407 &  	 3872 & Potsdam & $-$67 & 72 & 0.9 & 44 & 88 & 0.5 & $-$28 & 59 & 0.5 & 49 & 71 & 0.7 \\ 	 & \multicolumn{1}{c}{12/555}	& 3779 & Bonn & 3 & 65 & 0.0 & 42 & 69 & 0.6 & 73 & 49 & 1.5 & 76 & 49 & 1.6 \\ 
HD\,164844 & 56810.41567 &  	 2897 & Potsdam & $-$148 & 65 & 2.3 & 4 & 78 & 0.1 & $-$72 & 51 & 1.4 & 67 & 59 & 1.1 \\ 	 & \multicolumn{1}{c}{8/1130}	& 2865 & Bonn & $-$180 & 61 & 3.0 & $-$28 & 61 & 0.5 & $-$86 & 41 & 2.1 & $-$14 & 42 & 0.3 \\ 
HD\,95568 & 56811.02373 &  	 3170 & Potsdam & 57 & 51 & 1.1 & $-$59 & 58 & 1.0 & 13 & 35 & 0.4 & $-$42 & 37 & 1.1 \\ 	 & \multicolumn{1}{c}{8/3050}	& 3104 & Bonn & 73 & 47 & 1.6 & $-$50 & 45 & 1.1 & 38 & 29 & 1.3 & $-$17 & 29 & 0.6 \\ 
HD\,156134 & 56811.07084 &  	 3789 & Potsdam & 28 & 51 & 0.5 & 72 & 63 & 1.1 & $-$46 & 26 & 1.8 & 25 & 31 & 0.8 \\ 	 & \multicolumn{1}{c}{12/2260}	& 3666 & Bonn & 38 & 53 & 0.7 & 35 & 48 & 0.7 & 9 & 32 & 0.3 & 8 & 31 & 0.3 \\ 
HD\,156292 & 56811.11206 &  	 3619    & Potsdam & $-$75 & 75 & 1.0 & 32 & 63 & 0.5 & 33 & 51 & 0.6 & 11 & 49 & 0.2 \\ 	 & \multicolumn{1}{c}{12/810}	& 3619 & Bonn & $-$24 & 55 & 0.4 & 41 & 47 & 0.9 & 24 & 35 & 0.7 & 18 & 34 & 0.5 \\ 
HD\,97991 & 56811.13721 &  	 3619 & Potsdam & $-$154 & 81 & 1.9 & 62 & 75 & 0.8 & $-$51 & 62 & 0.8 & 41 & 56 & 0.7 \\ 	 & \multicolumn{1}{c}{12/1110}	& 3496 & Bonn & $-$106 & 72 & 1.5 & 15 & 66 & 0.2 & 48 & 52 & 0.9 & $-$46 & 50 & 0.9 \\ 
HD\,156234 & 56812.15657 &  	 3819 & Potsdam & $-$95 & 50 & 1.9 & 4 & 88 & 0.0 & $-$114 & 31 & 3.7 & 10 & 50 & 0.2 \\ 	 & \multicolumn{1}{c}{10/2350}	& 3813 & Bonn & $-$53 & 62 & 0.9 & $-$26 & 56 & 0.5 & $-$61 & 43 & 1.4 & 4 & 38 & 0.1 \\ 
{\bf HD\,54879} & 57099.00901 &  	 3060 & Potsdam & $-$633 & 65 & 9.7 & 33 & 68 & 0.5 & $-$527 & 45 & 11.7 & 52 & 45 & 1.2 \\ 	 & \multicolumn{1}{c}{8/545}	& 2925 & Bonn & $-$600 & 79 & 7.6 & 1 & 58 & 0.0 & $-$471 & 39 & 12.1 & 69 & 35 & 2.0 \\ 
HD\,56779 & 57099.02950 &  	 3363 & Potsdam & $-$84 & 42 & 2.0 & 48 & 57 & 0.8 & $-$61 & 33 & 1.8 & 3 & 41 & 0.1 \\ 	 & \multicolumn{1}{c}{10/60}	& 3058 & Bonn & $-$121 & 37 & 3.3 & $-$2 & 32 & 0.1 & $-$72 & 29 & 2.5 & $-$4 & 25 & 0.2 \\ 
HD\,75759 & 57099.05031 &  	 2721 & Potsdam & $-$102 & 71 & 1.4 & 57 & 75 & 0.8 & $-$103 & 58 & 1.8 & $-$34 & 77 & 0.4 \\ 	 & \multicolumn{1}{c}{8/120}	& 2936 & Bonn & $-$32 & 63 & 0.5 & 54 & 57 & 0.9 & $-$1 & 45 & 0.0 & $-$41 & 41 & 1.0 \\ 
HD\,110432 & 57099.13247 &  	 4708 & Potsdam & $-$55 & 48 & 1.1 & 55 & 55 & 1.0 & $-$62 & 39 & 1.6 & 66 & 48 & 1.4 \\ 	 & \multicolumn{1}{c}{18/244}	& 4606 & Bonn & 544 & 200 & 2.7 & 129 & 183 & 0.7 & 38 & 66 & 0.6 & 39 & 58 & 0.7 \\ 
HD\,120324 & 57099.16026 &  	 4833 & Potsdam & 9 & 40 & 0.2 & 80 & 48 & 1.7 & $-$22 & 31 & 0.7 & 4 & 35 & 0.1 \\ 	 & \multicolumn{1}{c}{14/28}	& 4362 & Bonn & 47 & 44 & 1.1 & 26 & 33 & 0.8 & 22 & 27 & 0.8 & 7 & 23 & 0.3 \\ 
HD\,120991 & 57099.18338 &  	 5919 & Potsdam & $-$76 & 35 & 2.2 & $-$15 & 39 & 0.4 & $-$85 & 26 & 3.3 & $-$21 & 28 & 0.8 \\ 	 & \multicolumn{1}{c}{18/745}	& 5595 & Bonn & $-$77 & 60 & 1.3 & $-$26 & 32 & 0.8 & $-$76 & 25 & 3.0 & $-$15 & 19 & 0.8 \\ 
{\bf CPD\,$-$62$^{\circ}$\,2124} & 57099.21385 &  	 1820 & Potsdam & 4636 & 128 & 36.2 & 28 & 102 & 0.3 & 4531 & 98 & 46.2 & 9 & 76 & 0.1 \\ 	 & \multicolumn{1}{c}{8/4800}	& 1767 & Bonn & 5222 & 123 & 42.5 & 27 & 95 & 0.3 & 4399 & 65 & 67.7 & $-$31 & 53 & 0.6 \\ 
HD\,156233 & 57099.28267 &  	 2222 & Potsdam & $-$92 & 124 & 0.7 & 34 & 117 & 0.3 & $-$139 & 76 & 1.8 & 19 & 79 & 0.2 \\ 	 & \multicolumn{1}{c}{8/1440}	& 2215 & Bonn & $-$34 & 102 & 0.3 & 86 & 92 & 0.9 & $-$50 & 68 & 0.7 & 48 & 63 & 0.8 \\ 
HD\,315032 & 57099.34159 &  	 3660 & Potsdam & $-$48 & 78 & 0.6 & $-$22 & 85 & 0.3 & $-$50 & 51 & 1.0 & $-$37 & 59 & 0.6 \\ 	 & \multicolumn{1}{c}{16/3780}	& 3581 & Bonn & $-$69 & 75 & 0.9 & $-$31 & 56 & 0.6 & $-$43 & 47 & 0.9 & 10 & 41 & 0.2 \\ 

\hline
\end{tabular}
\end{table*}

\begin{figure*}
\centering
\includegraphics[width=.35\textwidth]{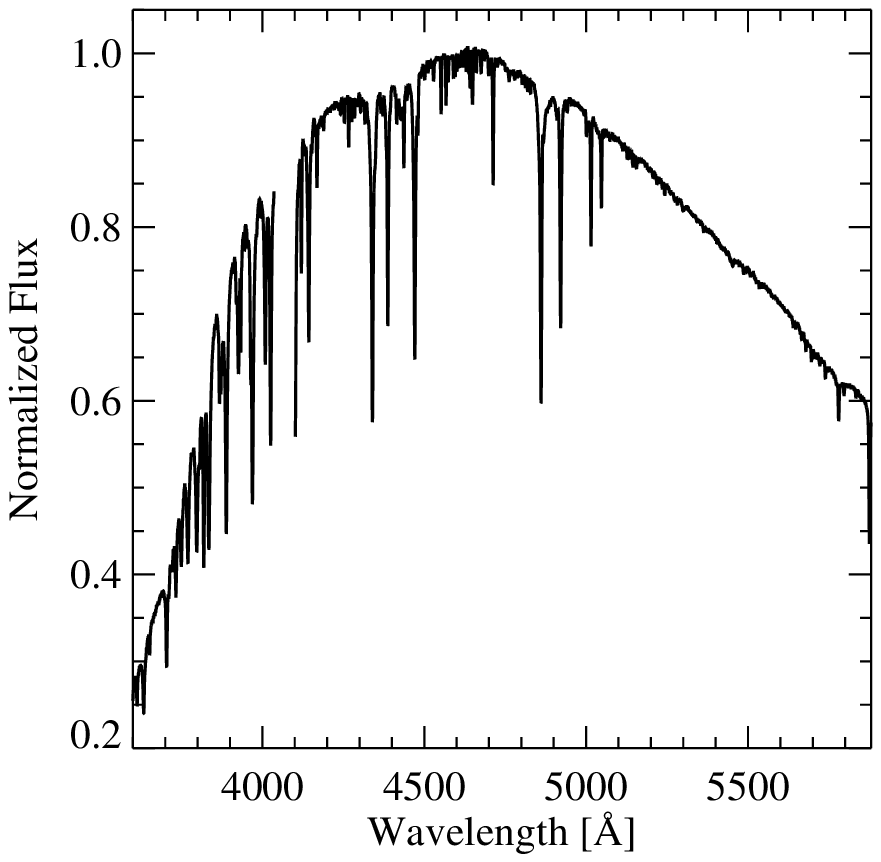}
\includegraphics[width=.35\textwidth]{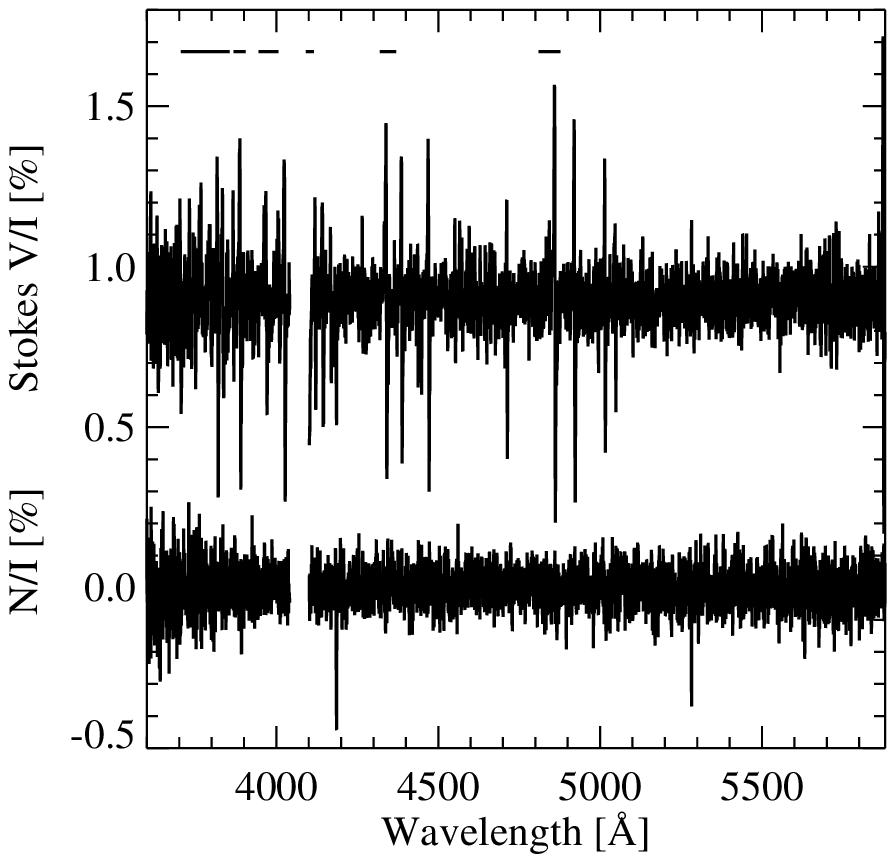}
\includegraphics[width=.30\textwidth]{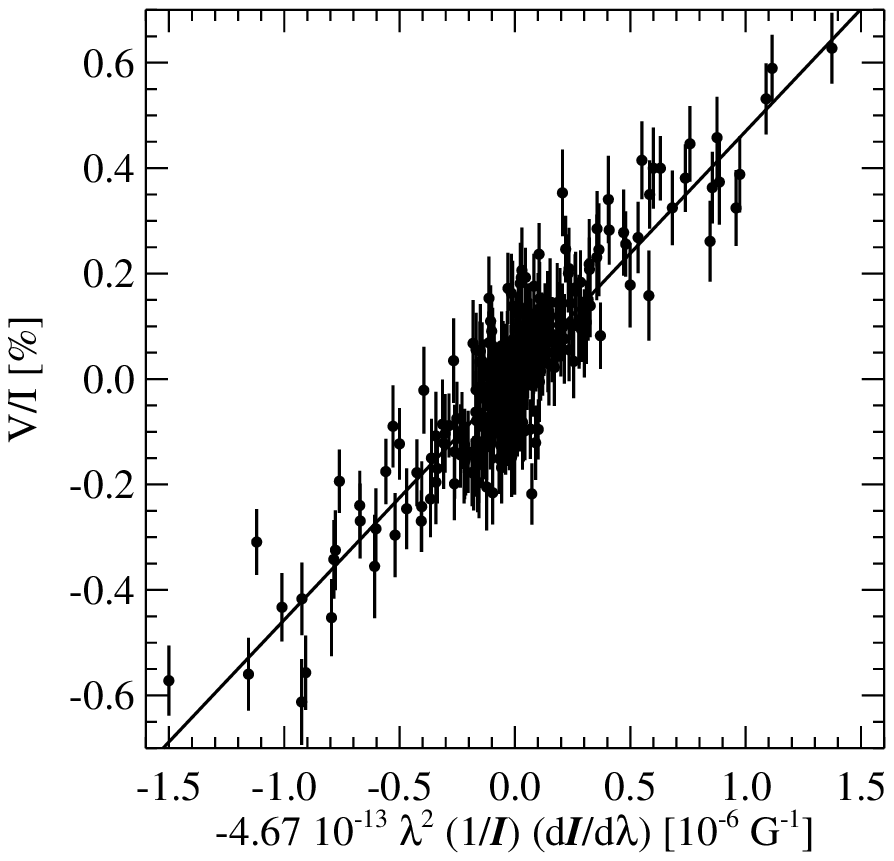}
\includegraphics[width=.30\textwidth]{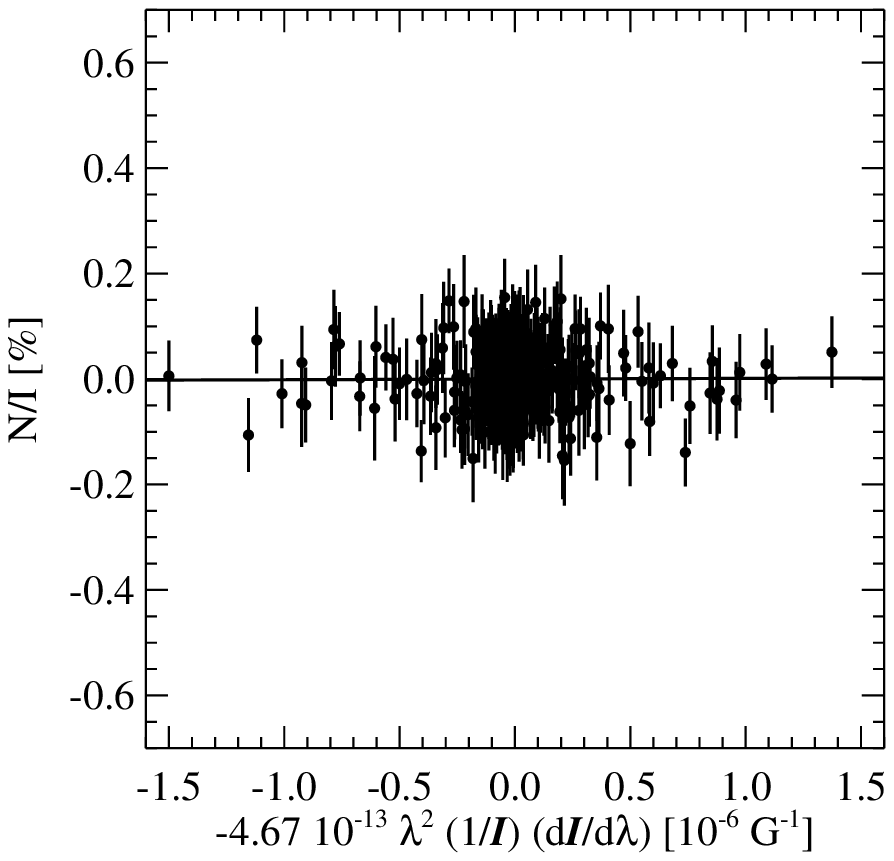}
\includegraphics[width=.30\textwidth]{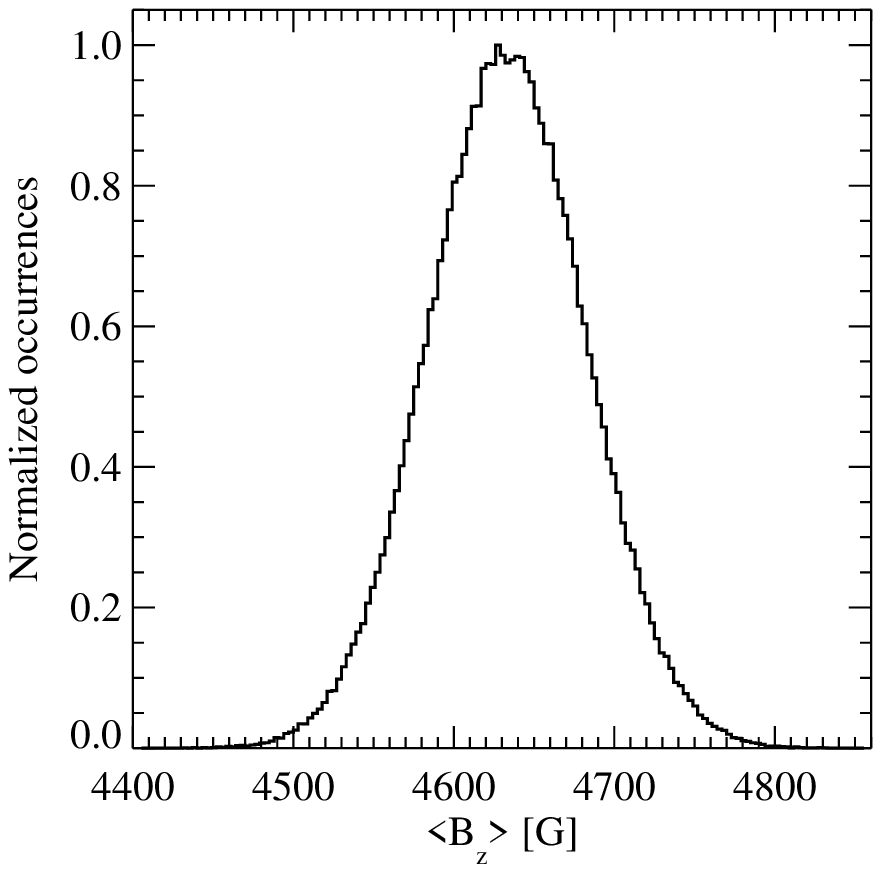}
\caption{
Overview of the results of the analysis of the FORS\,2 data of CPD\,$-$62$^{\circ}$\,2124,
collected during the night of 2015 March 17, considering the hydrogen lines,
using the Potsdam pipeline.
{\sl Top left panel:}
observed Stokes~$I$ spectrum arbitrarily normalized to the highest value.
{\sl Top right panel:}
the top profile shows Stokes~$V$ (in \%), while the bottom profile shows the
spectrum of the $N$ parameter (in \%).
The Stokes~$V$ spectrum is shifted upwards by 0.9\% for better visibility.
The regions used to calculate the longitudinal magnetic field are marked by
horizontal lines close to the top of the panel.
{\sl Bottom left panel:}
linear fit to Stokes~$V$.
{\sl Bottom middle panel:}
linear fit to the $N$ spectrum.
From the linear fit, we determine $\left<N_{\rm z}\right> = 28\pm102$\,G.
{\sl Bottom right panel:}
distribution of the longitudinal magnetic field values $P(\left<B_{\rm z}\right>)$,
obtained via bootstrapping.
From the distribution $P(\left<B_{\rm z}\right>)$, we obtain the most likely value for
the longitudinal magnetic field $\left<B_{\rm z}\right> = 4636\pm128$\,G.
We note that the gaps in the region around H$\delta$ in the two upper panels result
from masking an internal reflection in that spectral range.
}
\label{fig:cpd-62}
\end{figure*}

We present the results of our magnetic field measurements in Table~\ref{tab:MFs}.
The first column gives the object name, followed by the modified Julian date,
as well as the number of exposures and the total exposure time in Col.~2,
and the peak value of the S/N for the extracted full Stokes~$I$ spectrum, separately for the two groups, in Col.~3.
The S/N is calculated per \AA{} as the median of the 100~pixels with the highest flux,
excluding emission lines.
In the next column, we list the reduction software used for the determination of
the magnetic field (``Bonn'' or ``Potsdam'').
Columns~5 and 9 show the values for the longitudinal magnetic field determined from
the Stokes~$V$ spectrum, for the hydrogen lines or for the whole spectrum, respectively.
Columns~7 and 11 give the same values, but determined from the $N$ spectrum.
Columns~6, 8, 10, and 12 indicate the significance of the respective measurements,
defined as the absolute value of the measurement divided by its 1$\sigma$ error.
Please note that the wavelength ranges used both for the hydrogen lines and for the whole spectrum
might differ between the two pipelines.
Object names are given in bold face, if a magnetic field was found in the measurement.

The detection of strong magnetic fields in the stars HD\,54879, CPD\,$-$57$^{\circ}$\,3509, and HD\,164492C
was already discussed in the previous paper on BOB FORS\,2 observations (Fossati et al.\ \cite{Fossati2015}).
Individual studies of HD\,54879, HD\,164492C, and CPD\,$-$57$^{\circ}$\,3509
were presented in the papers by Castro et al.\ (\cite{Castro2015}),
Hubrig et al.\ (\cite{Hubrig2014a}), and Przybilla et al.\ (\cite{Przybilla2016}), respectively.
We have reobserved these objects and find rotational modulation of the magnetic fields
in HD\,164492C and CPD\,$-$57$^{\circ}$\,3509, and confirm the magnetic field in HD\,54879.

The magnetic nature of the star HD\,345439 was already discussed in the work of Hubrig et al.\ (\cite{Hubrig2015a}).
The analysis of HD\,345439 using four subsequent spectropolarimetric FORS\,2 subexposures does not reveal a
magnetic field.
On the other hand, Hubrig et al.\ (\cite{Hubrig2015a}) report that the
individual subexposures indicate that
HD\,345439 may host a strong magnetic field that rapidly varies over 88\,min.
The fast rotation of HD\,345439 is also
indicated by the behavior of several metallic and \ion{He}{i} lines in the low-resolution FORS\,2 spectra that
show profile variations already on this short time-scale.
Wisniewski et al.\ (\cite{Wisniewski2015}) found 
clear evidence that the strength of H$\alpha$, \ion{He}{i}, and the Brackett series lines
indeed vary on a time scale of $\sim$0.7701\,d from their analysis of multi-epoch,
multi-wavelength spectroscopic monitoring.

Apart from these detections, a very strong mean longitudinal magnetic field
$\left<B_{\rm z}\right> = 5222\pm123$\,G (using the Bonn pipeline on the hydrogen lines) in the
rather faint ($V=11.0$) early B-type star CPD\,$-$62$^{\circ}$\,2124 is discovered in our survey.
In Fig.~\ref{fig:cpd-62}, we illustrate the analysis of the hydrogen lines for this object,
using the Potsdam pipeline.
Such a strong longitudinal magnetic field implies
a dipolar magnetic field strength of more than 17\,kG.
Massive stars with such extremely strong magnetic fields are very rare
(Bychkov et al.\ \cite{Bychkov2009}; Petit et al.\ \cite{Petit2013}; Fossati et al.\ \cite{Fossati2015}).
Follow-up measurements of CPD\,$-$62$^{\circ}$\,2124 using HARPSpol
on one occasion also confirm the presence of an extraordinarily strong magnetic field.
A paper presenting an
individual study of this extremely interesting star was recently submitted by Castro et al.

Another interesting result is achieved for the star BD\,$-$12\,4982,
where the magnetic field is measured at the 4.7$\sigma$ significance level using the Bonn pipeline and
5.2$\sigma$ using the pipeline in Potsdam.
These results are achieved when the whole spectrum is used for the measurements.
The measurements using only the hydrogen lines show a significance level below 3$\sigma$.
Work on this object using HARPS observations is currently ongoing
(J\"arvinen et al., in preparation).

For a few stars, HD\,56779, HD\,120991, CD\,$-$22$^{\circ}$\,12513, and HD\,164844, longitudinal magnetic fields are
detected at a significance level between 3--3.3$\sigma$,
using either both reductions from Bonn and Potsdam or only one of the pipelines.
It will be important to monitor these stars with additional
FORS\,2 observations or to obtain high-resolution
spectropolarimetric observations to scrutinize more methodically for the presence of a magnetic field.

It would be especially interesting to carry out a follow-up study of the Be star HD\,120991,
since there is currently no undisputed evidence of a magnetic field in any classical Be star.
Hubrig et al.\ (\cite{Hubrig2009}) carried out a search for
magnetic fields in 16~Be~stars and concluded that magnetic fields in such stars are usually very weak, below 100\,G.
Bagnulo et al.\ (\cite{Bagnulo2012,Bagnulo2015}) could not confirm the magnetic field detections from the same data sets.
The MiMeS group reported that none of the 85 Be stars studied with ESPaDOnS, Narval, and HARPSpol
showed the presence of a magnetic field, with a median 1$\sigma$ error of 103\,G (Wade et al.\ \cite{Wade2014}).
In our sample of 71~stars, we included four Be stars, HD\,110432, HD\,120324, HD\,120991, and
HD\,315032.
For the first three Be stars, hard X-ray emission was detected using {\em XMM-Newton} observations (see Sect.~\ref{sect:xrays}).
Only for the Be star HD\,120991 do we obtain magnetic fields at a 3--3.3$\sigma$ level,
using both pipelines applied to the whole spectrum.
Should the presence of a magnetic field in HD\,120991 be established with additional measurements,
it would be weak.

\section{In depth analysis of the magnetic field measurements}
\label{sect:analysis}

\subsection{Comparison of the two data reduction pipelines}
\label{sect:comparison}

The FORS\,2 spectropolarimetric data obtained within the BOB collaboration
was independently reduced and analyzed by two teams using different tools and pipelines.
This gives us the possibility to directly compare the results
for a statistically large sample of stars.

\begin{figure*}
\centering
\includegraphics[width=.40\textwidth]{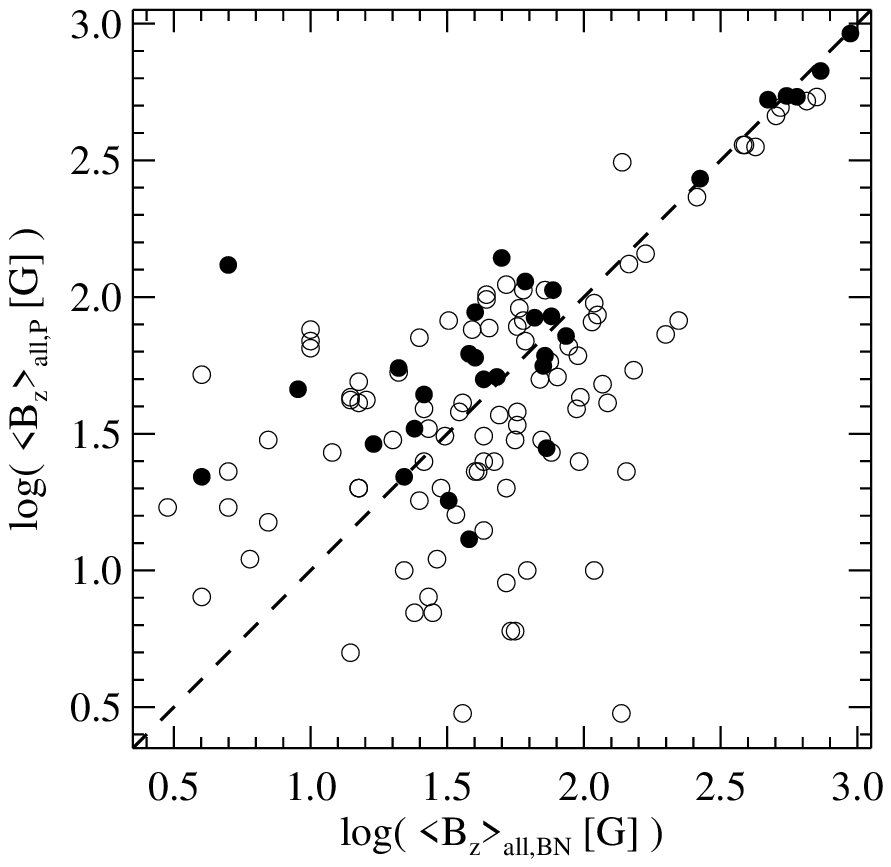}
\includegraphics[width=.40\textwidth]{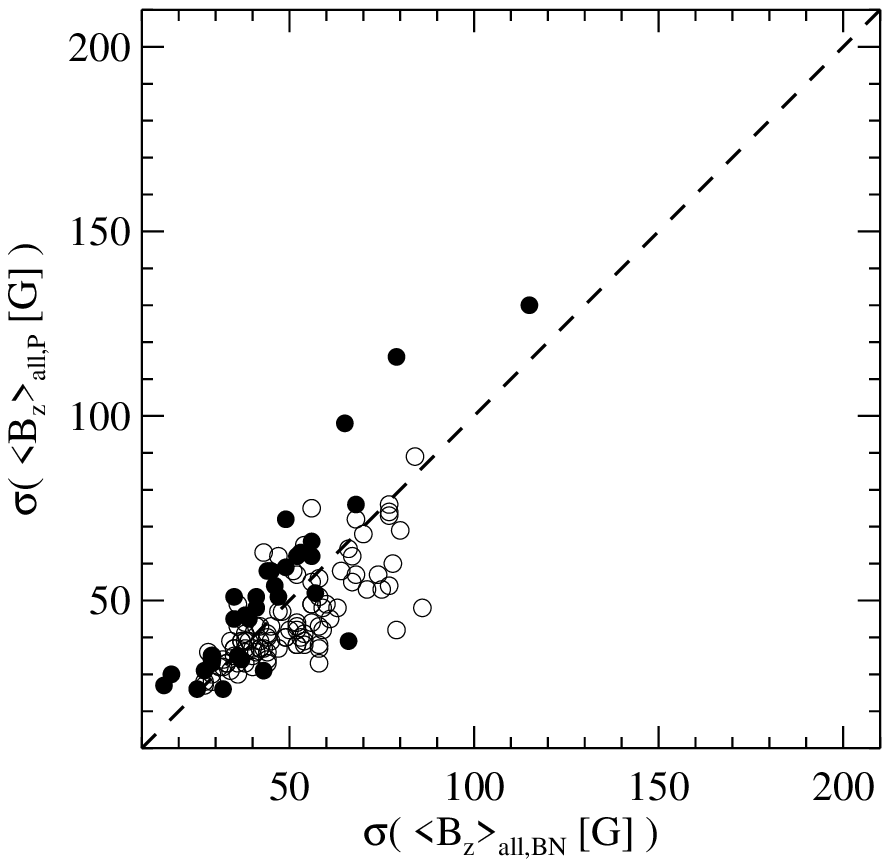}
\includegraphics[width=.40\textwidth]{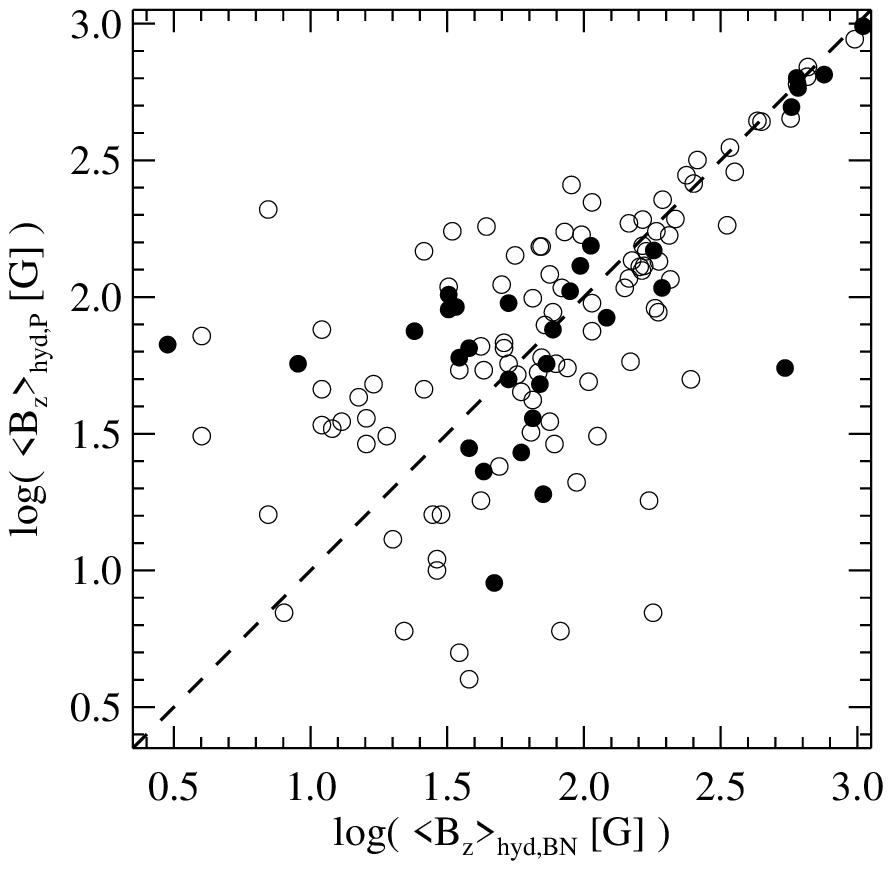}
\includegraphics[width=.40\textwidth]{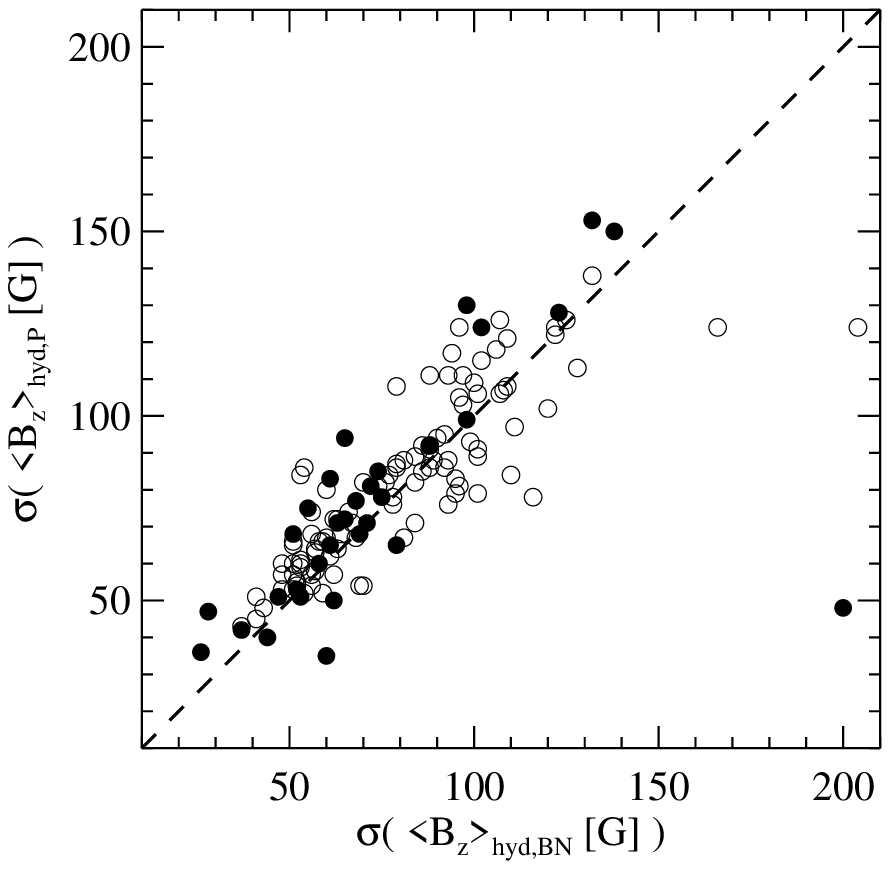}
\caption{
{\sl Top left panel:}
comparison between the $|\left<B_{\rm z}\right>|$ values obtained by analyzing
the whole spectrum with the Bonn pipeline and the Potsdam pipeline.
{\sl Top right panel:}
same as top left panel, but for the uncertainties of the $\left<B_{\rm z}\right>$ values
obtained analyzing the whole spectrum.
{\sl Bottom left panel:}
same as top left panel, but for the $|\left<B_{\rm z}\right>|$ values obtained
analyzing only the hydrogen lines.
{\sl Bottom right panel:}
same as top left panel, but for the uncertainties of the $\left<B_{\rm z}\right>$ values
obtained analyzing only the hydrogen lines.
Filled circles are from measurements presented in this paper, while open circles
are from the measurements presented in Fossati et al.\ (\cite{Fossati2015}).
The extreme outlier in the bottom right panel in the lower right is from the Be star
HD\,110432 (see the discussion in the text).
}
\label{fig:comparisons}
\end{figure*}

Figure~\ref{fig:comparisons} shows the comparison between the results obtained
by reducing and analysing the spectra
with the Bonn and Potsdam pipelines.
We consider here all 134 sets of measurements (this work and Fossati et al.\ \cite{Fossati2015}),
each set composed of two measurements
(i.e., $\left<B_{\rm z}\right>$ obtained from the analysis of the hydrogen lines or of the whole spectrum),
and obtained by the two pipelines, for a total of 536 measurements.
As was already evident from the analysis of a subset of the data we present in Fig.~\ref{fig:comparisons}
(see Fossati et al.\ \cite{Fossati2015}), there is a general good agreement between the reduction and analysis
of the two groups, with the occasional outlier.

To allow a better quantification of the agreement between the results of the two groups,
we have compared the difference of the two distributions of magnetic field measurements from
Bonn (BN) and Potsdam (P) with a Gaussian distribution.
For this, we computed for each of the 134 measurements the following value:

\begin{equation}
\left<B_{\rm z}\right>_{\rm diff} = \frac{ \left<B_{\rm z}\right>_{\rm BN} - \left<B_{\rm z}\right>_{\rm P} }
   {\sqrt{\sigma_{\left<B_{\rm z}\right>,{\rm BN}}^2 + \sigma_{\left<B_{\rm z}\right>,{\rm P}}^2}},
\label{eqn:err_norm_diff}
\end{equation}

\noindent
which is the error-normalized difference between the $\left<B_{\rm z}\right>$ values measured
by both the Bonn and Potsdam groups.
The normalization was made using the square root of the sum of the squared errors coming from the
Bonn and Potsdam analyses, following error propagation.

\begin{figure}
\centering
\includegraphics[width=.35\textwidth]{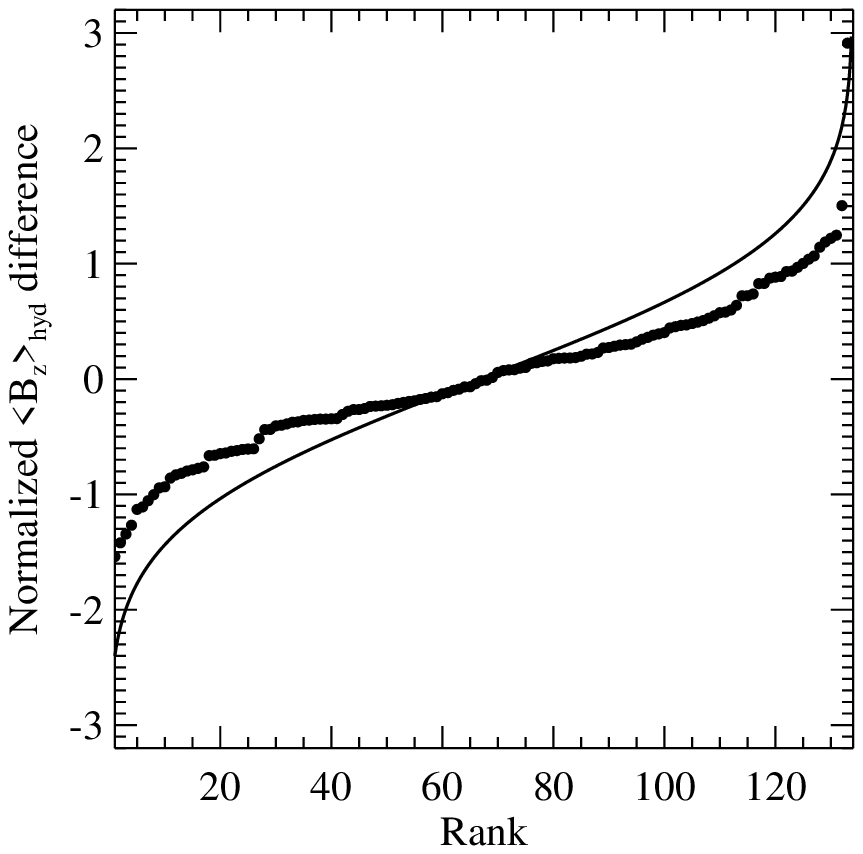}
\includegraphics[width=.35\textwidth]{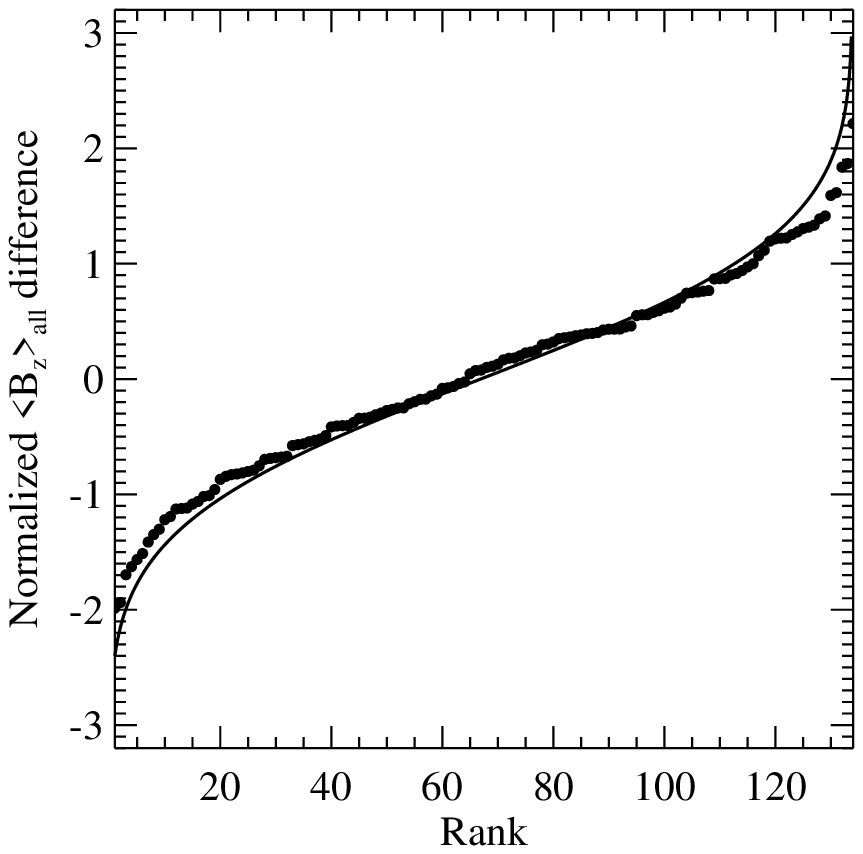}
\caption{
Density distributions for the error-normalized differences between the magnetic field values
determined by the Bonn and Potsdam groups for the 134 measurements presented in Fossati et al.\ (\cite{Fossati2015})
and in this paper.
The overlayed lines are for a Gaussian density distribution.
{\em Top panel:}
Normalized differences for the $\left<B_{\rm z}\right>$ obtained from hydrogen lines.
{\em Bottom panel:}
Normalized differences for the $\left<B_{\rm z}\right>$ obtained from all lines.
}
\label{fig:density_distribution}
\end{figure}

In Fig.~\ref{fig:density_distribution}, we show the resulting density distributions for the
differences from the measurements of the hydrogen lines (top) and from the whole spectrum (bottom).
In these graphs, all values calculated according to Eq.~(\ref{eqn:err_norm_diff})
are sorted and then plotted corresponding to their rank in the distribution.
The solid overlayed line shows the density distribution for a Gaussian.
While the density distribution for the normalized differences for the $\left<B_{\rm z}\right>$ obtained from all lines
corresponds well with a Gaussian distribution, the equivalent density distribution for the hydrogen lines
is much narrower, owing to the larger errors.
We would like to stress that strictly speaking it is only possible to make statistical comparisons
in the way presented here for repeated measurements of the same value,
but not for individual measurements of different values.
Also, the difference of two similar Gaussian distributions results in a distribution that is
a factor of $\sqrt{2}$ wider than the original distributions, i.e.\ the distribution coming
from the difference of our two measurements is narrower than should be expected,
pointing to a correlation between the Bonn and Potsdam reduction and analysis
approaches.

Please note that one measurement is not shown for the hydrogen density distribution, since the 
difference between the measurements is quite large.
This outlier is CPD\,$-$62$^{\circ}$\,2124, where we have a normalized difference of 3.3.
This is due to the 12\% difference in the value for the magnetic field determined
by both groups, and the high significance of the result (see also Fossati et al.\ \cite{Fossati2015}).

Overall, our results show that differences between the two
data reduction and analysis pipelines are usually well within 3$\sigma$.
The differences become important 
when evaluating measurements near the 3$\sigma$ threshold and when assessing the
absolute values of the magnetic fields measured with high significance.

\subsection{S/N in spectra and resulting errors}
\label{sect:SNR}

\begin{figure}
\centering
\includegraphics[width=.35\textwidth]{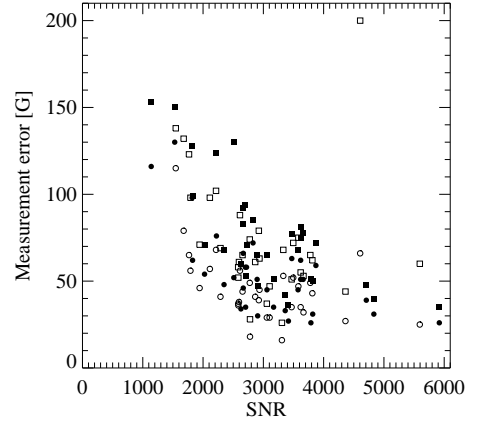}
\caption{
Derived 1$\sigma$ errors plotted against the obtained S/N for the
magnetic field measurements presented in Table~\ref{tab:MFs}.
Circles denote measurements using the whole spectrum,
squares measurements using only the hydrogen lines,
filled symbols results from the Potsdam reduction and analysis,
and open symbols results from the Bonn reduction and analysis.
}
\label{fig:snr}
\end{figure}

In Fig.~\ref{fig:snr}, we show the dependence of the derived errors of
the magnetic field measurements on the S/N obtained in the
raw spectra.
It is obvious that with higher S/N, i.e.\ higher flux,  the errors become smaller.
Using the whole spectrum (circles) results in smaller errors than using only
the hydrogen lines (squares).
Also, the Potsdam reduction usually delivers spectra with a slightly higher S/N
than the Bonn reduction.
The errors derived from the Bonn analysis are usually slightly smaller than from the Potsdam
analysis, mainly due to accounting for deviations from the nominal CCD gain
via $\chi^2$-scaling (see Fossati et al.\ \cite{Fossati2015}).
The outlier from the Bonn pipeline using only the hydrogen lines is from HD\,110432,
where the Bonn pipeline excludes the various emission lines.

The measurement error $\sigma$ depends not only on the S/N,
but also on the nature of the spectrum,
i.e.\ spectral type, $v\,\sin\,i$, binarity, etc.
For the same S/N, $\sigma$ is expected to be lower for observations
of early B-type stars compared to observations of O-type stars.
FORS\,2 magnetic field detections in O-type stars
that have been confirmed with high-resolution spectropolarimetry rarely reach
4$\sigma$.
Based on our data, where only very few stars were studied at a S/N higher than 4000,
we can not make conclusions on the saturation of $\sigma$ at even higher S/N.

Bagnulo et al.\ (\cite{Bagnulo2015}) presented a comparable analysis (see their Fig.~5) for
$\sim$1400 data sets obtained with FORS\,1 in spectropolarimetric mode, mainly concentrating on the
errors obtained from the analysis of the null spectra.
While their results are qualitatively similar, it can be seen that our data typically have
higher S/N than the heterogeneous FORS\,1 sample.

\subsection{Rectification}
\label{sect:rectification}

All Stokes~$V$ spectra were rectified
to ensure that the continuum is consistent with zero
(see as an example the work by
Hubrig et al.\ \cite{Hubrig2014a} and Fossati et al.\ \cite{Fossati2015}).
The offsets of the non-rectified Stokes~$V$ spectra from 0 are typically small, below 0.0015.
Rectification usually leads to a vertical shift of the spectra.
Among the whole sample of 71~stars discussed in this paper and in Fossati et al.\ (\cite{Fossati2015}),
the non-rectified $V$~spectrum is tilted only for the Be star HD\,110432
and the spectroscopic binary HD\,92206C, with an O6.5V((f)) primary.
Such a behavior of the Stokes~$V$ spectrum is sometimes observed in Herbig Ae/Be stars
due to the presence of a circumstellar disk (Hubrig et al., in preparation).
A variable tilt of the Stokes~$V$ spectrum over the orbital phase was detected in the Cyg\,X-1 system,
which consists of an O-type supergiant and a black hole (Karitskaya et al.\ \cite{Karitskaya2010}).
Usually, this type of feature is attributed to cross-talk between linear and circular polarization
and is seen in sources that show strong linear polarization
(e.g.\
Bochkarev \& Karitskaya \cite{BochkarevKaritskaya2012};
Bagnulo et al. \cite{Bagnulo2012}).

\begin{figure}
\centering
\includegraphics[width=.24\textwidth]{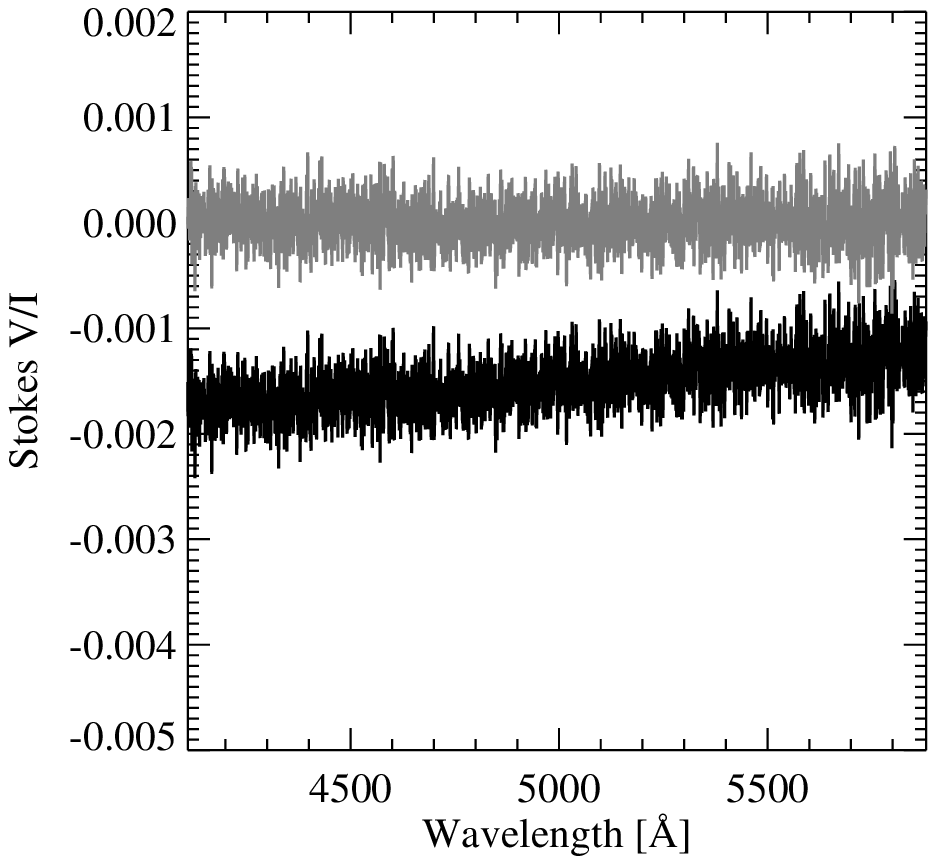}
\includegraphics[width=.24\textwidth]{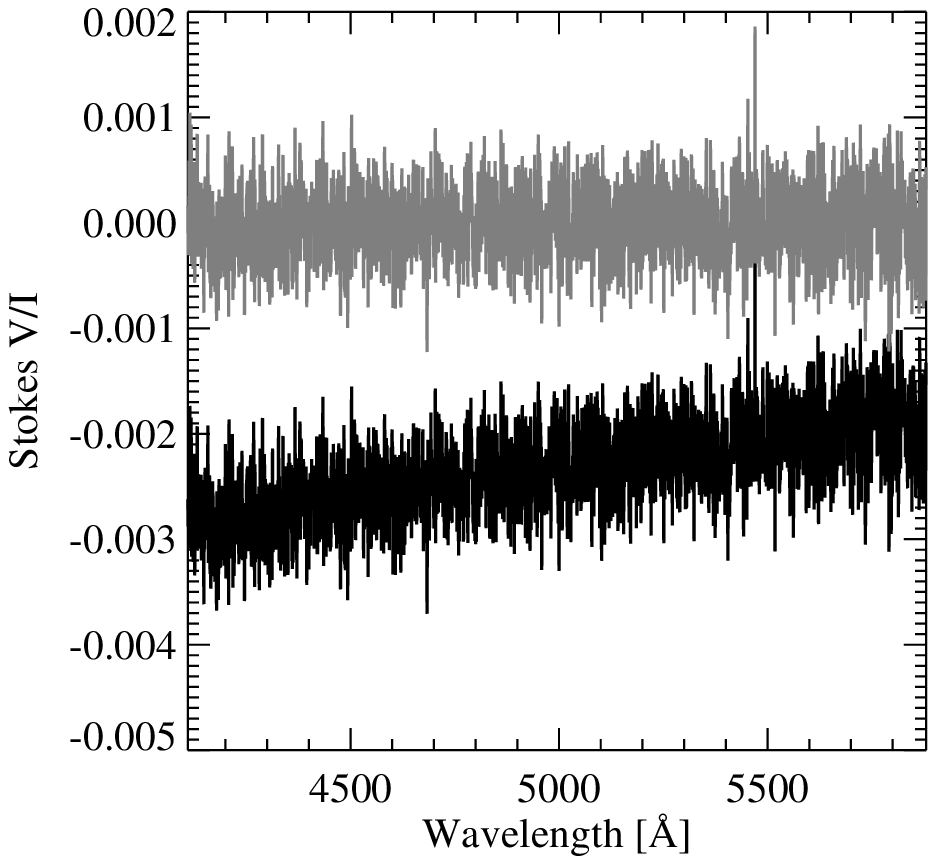}
\includegraphics[width=.24\textwidth]{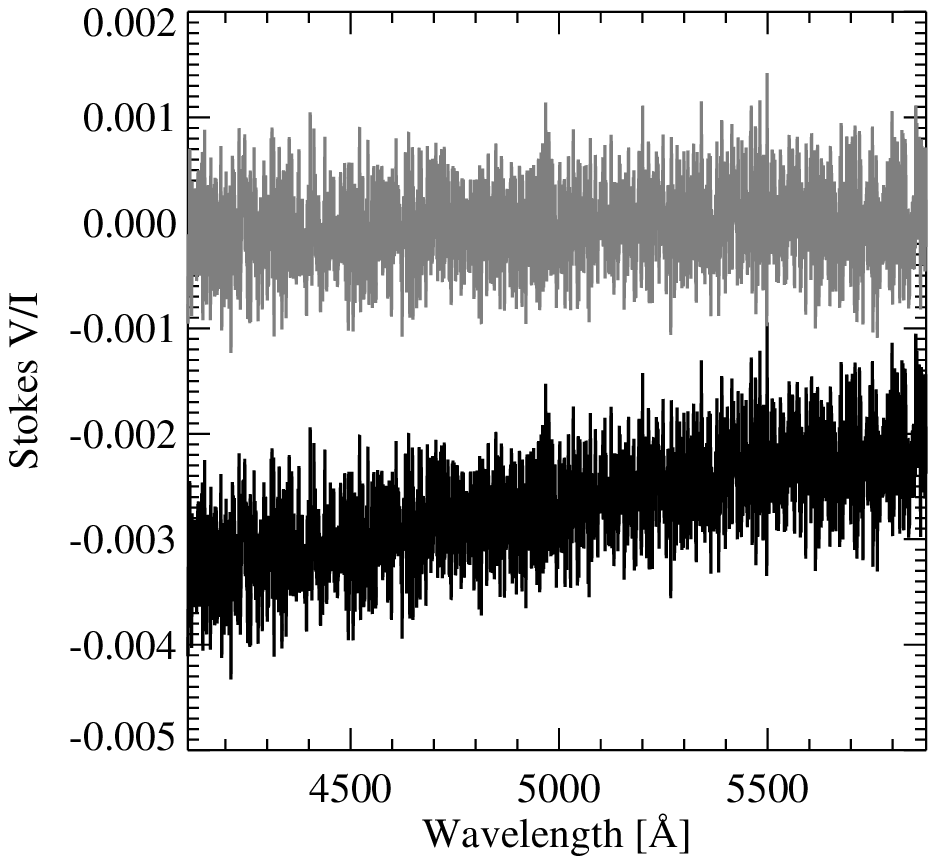}
\includegraphics[width=.24\textwidth]{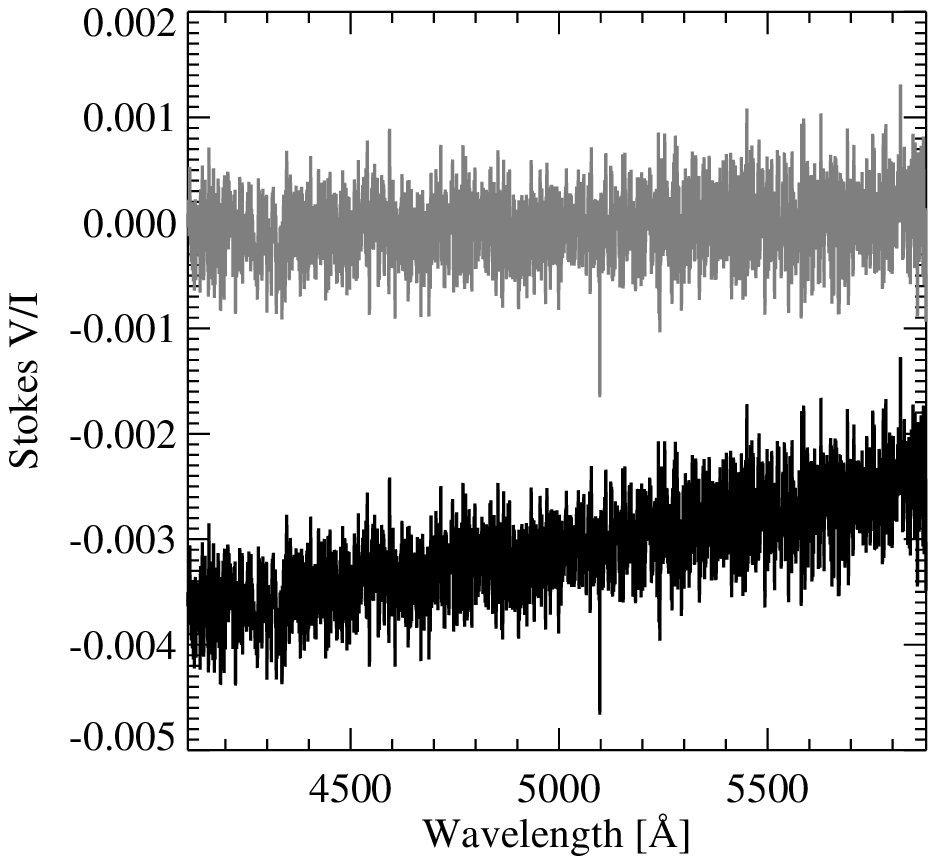}
\caption{
Non-rectified (black) and rectified (grey) Stokes $V/I$ spectra
of the Be star HD\,110432 (top left) and
the spectroscopic binary HD\,92206C with an O6.5V((f)) primary on three consecutive nights,
determined using the whole spectrum and the Potsdam pipeline.
}
\label{fig:spectrum_rect}
\end{figure}

In Fig.~\ref{fig:spectrum_rect}, we present the rectified and non-rectified spectra of HD\,110432 and
for the three observations of HD\,92206C.
While the rectification has only some influence on the determination of the magnetic field
in HD\,110432, leading to a value of $\left<B_{\rm z}\right> = 22\pm46$\,G for the magnetic field from
the non-rectified spectrum compared to $\left<B_{\rm z}\right> = -62\pm39$\,G from the rectified spectrum,
the influence is significant for HD\,92206C, where we obtain values for
the magnetic field of
$\left<B_{\rm z}\right> = 366\pm103$\,G,
$\left<B_{\rm z}\right> = 208\pm113$\,G, and
$\left<B_{\rm z}\right> = 239\pm94$\,G, respectively for the three nights from the non-rectified spectra,
compared to 
$\left<B_{\rm z}\right> = 76\pm64$\,G,
$\left<B_{\rm z}\right> = -48\pm72$\,G, and
$\left<B_{\rm z}\right> = -39\pm55$\,G from the rectified spectra.
All values were determined using the whole spectrum and
the Potsdam pipeline, which employs a linear fit to rectify the $V$ spectrum.

The Bonn pipeline, which uses a fourth-order polynomial for the rectification, leads to similar results.
Without rectification, the measurements of HD\,92206C give 
$\left<B_{\rm z}\right> = 357\pm98$\,G,
$\left<B_{\rm z}\right> = 209\pm104$\,G, and
$\left<B_{\rm z}\right> = 313\pm97$\,G, respectively,
while the rectified spectra lead to
$\left<B_{\rm z}\right> = 39\pm66$\,G,
$\left<B_{\rm z}\right> = -117\pm68$\,G, and
$\left<B_{\rm z}\right> = -26\pm56$\,G.

\subsection{Spectral variations due to pulsations}
\label{sect:pulsation}

A number of stars in our sample of 71 objects (this paper and Fossati et al.\ \cite{Fossati2015})
observed with FORS\,2
are reported in the literature to exhibit different types of pulsations.
From high-resolution spectropolarimetric observations, pulsations are known to have an impact on the
analysis of the presence of a magnetic field and its strength (e.g., Schnerr et al.\ \cite{Schnerr2006};
Hubrig et al.\ \cite{Hubrig2011b}; Neiner et al.\ \cite{Neiner2012}). 
Sometimes, it is also possible to observe features similar to Zeeman patterns expected in Stokes~$V$  spectra
in null spectra.
This depends on the pulsation phase, i.e.\ on the radial velocity
amplitude and the profile shape changes, and on the ratio between the exposure time and the pulsation period.
These features can be seen in observations
where the duration of the subexposures taken at different retarder waveplate angles
is a significant fraction of the pulsation cycle.
Unfortunately, so far the effect of pulsations on the low-resolution spectropolarimetric
observations with FORS\,2 has not been thoroughly investigated.

The following objects from the full sample of 71 stars are known to show pulsations.

{\it HD\,44597:}
This is a candidate short-period variable observed by Hipparcos and
found by Koen \& Eyer (\cite{KoenEyer2002}).
No frequency was found by Pigulski \& Pojmanski (\cite{PigulskiPojmanski2008}).
Buysschaert et al.\ (\cite{Buysschaert2015}) did not find a clear periodic variability with isolated frequencies
from {\em Kepler} data, but several low-frequency peaks stand out of the noise level.
The current speculative interpretation are convectively-driven internal gravity waves
(e.g.\ Rogers et al.\ \cite{Rogers2013})

{\it HD\,46149:}
Degroote et al.\ (\cite{Degroote2010}) detected
rotational modulation and stochastic $p$ modes from 3.05 to 7.23\,d$^{-1}$
from CoRoT data.

{\it HD\,46150 and HD\,46966:}
Blomme et al.\ (\cite{Blomme2011}) detected in both stars a ``red noise power excess'' in  CoRoT data, which was
interpreted as convectively-driven internal gravity waves of a stochastic nature
by Aerts \& Rogers (\cite{AertsRogers2015}).

{\it HD\,46202:}
Briquet et al.\ (\cite{Briquet2011}) used CoRoT data to detect
heat-driven p modes, similar to $\beta$\,Cephei-like modes with pulsation frequencies of $0.5-4.9$\,d$^{-1}$.

{\it HD\,46223:}
Blomme et al.\ (\cite{Blomme2011}) detected a ``red noise power excess'' in  CoRoT data, which was
interpreted as convectively-driven internal gravity waves of a stochastic nature
by Aerts \& Rogers (\cite{AertsRogers2015}).
The detected frequencies were 2.007, 4.011, and 13.792\,d$^{-1}$.

{\it HD\,46328:}
This is a well-known $\beta$\,Cephei star with a period of $\sim$5\,h
(Saesen et al.\ \cite{Saesen2006} and references therein).

{\it HD\,64365:}
Telting et al.\ (\cite{Telting2006}) found $\beta$\,Cephei pulsations of low degree.

{\it HD\,93521:}
Howarth \& Reid (\cite{HowarthReid1993}) and Rauw et al.\ (\cite{Rauw2012}) found that the stellar pulsations
in this star display a bi-periodic (1.75 and 2.89\,h) absorption line profile variability that is commonly
interpreted as the signature of two non-radial pulsation modes.

{\it HD\,95568:}
Pigulski \& Pojmanski (\cite{PigulskiPojmanski2008}) found $\beta$\,Cephei pulsations with a frequency
of 6.152355\,d$^{-1}$.

{\it HD\,117357:}
Pigulski \& Pojmanski (\cite{PigulskiPojmanski2008}) found $\beta$\,Cephei pulsations with frequencies of
2.08106(3)\,d$^{-1}$ and 6.53990\,d$^{-1}$.

{\it HD\,144470:}
Telting et al.\ (\cite{Telting2006}) found high degree $\beta$\,Cephei pulsations.

Not all stars that exhibit pulsations also showed changes between the spectra of the different subexposures.
HD\,46328, HD\,64365, and HD\,95568 were all observed three times, and
HD\,117357 was observed twice.
All spectra of these four stars showed profile changes within the observing sequence.
The single data set for HD\,46966 also showed small changes in the spectra.
HD\,44597 was observed in two nights.
While profiles changed during the first night, no variability was found in the second night.
HD\,46202 was observed three times.
Profile changes can be seen in the first two data sets, but there are
no changes in the last one.
No variations were found in the spectra of
HD\,46149, HD\,46150, and HD\,46223, each observed during a single epoch,
during the six epochs for HD\,144470, or the three epochs for HD\,93521.

It is possible that pulsations are a cause for spurious magnetic field detections.
However, they might also be traceable as significant values of the diagnostic $\left<N_{\rm z}\right>$.
Looking at the $\left<N_{\rm z}\right>$ measurements in the full data set of 71 stars,
there are five objects where $\left<N_{\rm z}\right>$ exceeds 3$\sigma$ in at least one of the measurements:
HD\,46328,
HD\,95568,
HD\,101008,
HD\,102475 (twice), and
HD\,289002.
While the first two objects are known pulsators (see above), nothing is known about pulsations in the
other three stars.
They all have a spectral type of B1 and might be pulsation candidates.

\section{Exploring indirect magnetic field indicators}
\label{sect:indicators}

Two properties were suggested in the past as indirect magnetic field indicators:
slow rotation and X-ray characteristics.
Furthermore, as was already reported in previous studies, the occurrence of magnetic stars
with radiative envelopes in close binary or multiple systems is extremely low
(e.g., Carrier et al.\ \cite{Carrier2002}; Sch\"oller et al.\ \cite{Schoeller2012};
Grunhut \& Wade \cite{GrunhutWade2013}).
This motivated Ferrario et al.\ (\cite{Ferrario2009}) to suggest a merger scenario to explain the
origin of the magnetic fields.
According to this scenario, if mergers were the sole mechanism to produce a magnetic
field in an OB star, no magnetic fields would be expected in a close binary.
To investigate the usefulness of these indicators,
we studied all 71 stars in our sample observed with FORS\,2.
About 1/3 of our targets were selected on the basis of their low $v\,\sin\,i$,
a few were explicitly selected for their hard X-rays.
In the following we discuss these properties for our FORS2 sample.

\subsection{Rotation}
\label{sect:vsini}

Morel et al.\ (\cite{Morel2008}) suggested that 
magnetic OB stars are mostly slowly rotating and nitrogen-rich.
Following their arguments, we selected 20 stars
that have a rotational velocity $v\,\sin\,i$ below 60\,km\,s$^{-1}$
from the catalogues of Howarth et al.\ (\cite{Howarth1997})
and Sim{\'o}n-D{\'{\i}}az \& Herrero (\cite{SimonDiazHerrero2014}).

From these 20 stars, only HD\,54879 shows a magnetic field.
The other three targets with detected magnetic fields were not selected on the basis of their low $v\,\sin\,i$.
In the subsequent studies
by Przybilla et al.\ (\cite{Przybilla2016}) on CPD\,$-$57$^{\circ}$\,3509
and Castro et al.\ (submitted) on CPD\,$-$62$^{\circ}$\,2124, both stars were
found to have a $v\,\sin\,i$ of 35\,km\,s$^{-1}$.
HD\,164492A was selected on the basis of its low $v\,\sin\,i$, but a magnetic field was found in the
component HD\,164492C (Hubrig et al.\ \cite{Hubrig2014a}; Gonz\'alez et al., submitted).

\subsection{Multiplicity}
\label{sect:binarity}

In the following, we list known companions to the systems in our sample.
We limited this compilation to companions closer than 3\arcsec{}, which corresponds
to the worst seeing conditions under which we observed.

{\it CPD\,$-$59$^{\circ}$\,2624:}
The Washington Double Star (WDS) Catalogue (Mason et al.\ \cite{Mason2001})
lists a companion at a distance of 0\farcs{}1 with a visual magnitude of 12.3.

{\it HD\,37020:}
This is $\theta^{1}$\,Ori\,A, which consists of two visual components.
The WDS Catalogue
gives a distance of 0\farcs{}2 for the pair.
Pourbaix et al.\ (\cite{Pourbaix2004}) list one of the two components
as a spectroscopic binary.
Its period is under discussion, with values of 65.4 or 6.5\,d.

{\it HD\,46150:}
The WDS Catalogue
lists companions at 2\farcs{}1 and 3\arcsec{}
with visual magnitudes of 13.65 and 12.4, respectively,

{\it HD\,46202:}
Sana et al.\ (\cite{Sana2014}) resolved a companion with the
Sparse Aperture Mask mode (SAM; Tuthill et al.\ \cite{Tuthill2010})
of NAOS-CONICA (Lenzen et al.\ \cite{Lenzen2003}; Rousset et al.\ \cite{Rousset2003})
at a distance of 85.5\,mas.

{\it HD\,46223:}
The WDS Catalogue
lists a companion at a distance of 0\farcs{}5 with a visual magnitude of 12.

{\it HD\,46966:}
The WDS Catalogue 
lists one companion at a distance of 0\farcs{}1 with a visual magnitude of 8.3.
Sana et al.\ (\cite{Sana2014}) detected this companion with SAM
at a distance of 50.5\,mas.

{\it HD\,72648:}
The WDS Catalogue 
lists one companion at a distance of 1\farcs{}7 with a visual magnitude of 14.2.

{\it HD\,72754:}
Pourbaix et al.\ (\cite{Pourbaix2004}) list this system as
a spectroscopic binary with a period of 33.734\,d.

{\it HD\,75759:}
Sota et al.\ (\cite{Sota2014}) list HD\,75759 as a double-lined spectroscopic binary.
Pourbaix et al.\ (\cite{Pourbaix2004})
give a period of 33.311\,d and an orbital eccentricity of 0.63.
Sana et al.\ (\cite{Sana2014}) resolved this system both in 2012 and 2013 with
PIONIER (Le~Bouquin et al.\ \cite{LeBouquin2011}) on the Very Large Telescope Interferometer
(e.g.\ Sch\"oller \cite{Schoeller2007}).
They found a sub-milliarcsecond companion, whose exact position and magnitude
in the H band were determined with very large error bars. 

{\it HD\,92206C:}
Sana et al.\ (\cite{Sana2014}) resolved this system with SAM
and found a companion at a distance of about 30\,mas.
Sota et al.\ (\cite{Sota2014}) list HD\,92206C as a double-lined spectroscopic binary.

{\it HD\,152218:}
Pourbaix et al.\ (\cite{Pourbaix2004}) list this object as a spectroscopic binary,
with a period of 5.4\,d.

{\it HD\,152246:}
Sana et al.\ (\cite{Sana2014}) resolved this system with PIONIER
and found a companion at a distance of 3\,mas.
Sota et al.\ (\cite{Sota2014}) list HD\,152246 as a double-lined spectroscopic binary.

{\it HD\,152590:}
Pourbaix et al.\ (\cite{Pourbaix2004}) list this object as a spectroscopic binary,
with a period of 4.5\,d.

{\it HD\,156292:}
This object is a double-lined spectroscopic binary with a period
of 4.94\,d (Sota et al.\ \cite{Sota2014}).

{\it HD\,164492A:}
Sana et al.\ (\cite{Sana2014}) resolved this system with both PIONIER and SAM
and found a companion at a distance of between 25 and 33\,mas.
The WDS Catalogue
lists another companion at a distance of 1\farcs{}5 with a visual magnitude of 13.
Sota et al.\ (\cite{Sota2014}) list HD\,164492A as a potential single-lined spectroscopic binary.

{\it HD\,164492C:}
Hubrig et al.\ (\cite{Hubrig2014a}) showed that at least two components are visible in the HARPS spectra
of this magnetic star.
Further analysis confirmed the presence of three components in the spectra (Gonz\'alez et al., submitted).
Note that according to the WDS Catalogue,
component HD\,164492D is at a distance of between 2\farcs{}3 and 2\farcs{}9.

{\it HD\,164536:}
The WDS Catalogue
lists a companion at distances between 1\farcs{}5 and 1\farcs{}7 with a visual magnitude of 12.4.

{\it HD\,164816:}
This system was resolved by Sana et al.\ (\cite{Sana2014}) both with PIONIER
and SAM in 2012.
PIONIER found the companion at a distance of 57\,mas with a magnitude difference of 3.5
in the H band.
Sota et al.\ (\cite{Sota2014}) list HD\,164816 as a double-lined spectroscopic binary.

{\it HD\,165052:}
Morrison \& Conti (\cite{MorrisonConti1978}) published the first orbit for this
spectroscopic binary with a period of 6.14\,d, which was later revised by
Stickland et al. (\cite{Stickland1997}) as 2.96\,d.

{\it HD\,168625:}
The WDS Catalogue
lists one companion at a distance of 1\farcs{}1 with a visual magnitude of 12.6.

In total, 16 out of the 71 objects in our sample have known companions at
a distance below 1\arcsec{}.
Among these multiple systems, a magnetic field at high confidence level was found only
in the star HD\,164492C.

\subsection{X-ray properties}
\label{sect:xrays}

\begin{table*}
\caption{Objects in our FORS\,2 sample with available X-ray observations.}
\label{tab:bstar}
\centering
\begin{tabular}{lccccc}
\hline
\hline
\multicolumn{1}{c}{Object} & 
\multicolumn{1}{c}{Spectral} &
\multicolumn{1}{c}{$L_{\rm X}$} &
\multicolumn{1}{c}{Adopted} &
\multicolumn{1}{c}{References} & 
\multicolumn{1}{c}{Comments} \\
 & 
\multicolumn{1}{c}{Type}&
 & 
\multicolumn{1}{c}{distance} & 
& 
\\
 & 
 & 
\multicolumn{1}{c}{[$10^{31}$\,erg\,s$^{-1}$]} &
\multicolumn{1}{c}{[kpc]} & 
 & 
\\
\hline
BD\,$-$12 4982 & B0\,II              & 1       & 1.5   & 1   & \\
HD\,110432     & B2\,pe              & 40      & 0.37  & 2,3 & $\gamma$\,Cas-analog \\
HD\,120324     & B2\,Vnpe            & 0.06    & 0.115 & 4   & \\
HD\,120991     & B2\,IIne            & 30      & 0.83  & 4,5 & $\gamma$\,Cas-analog ? \\
HD\,164816     & O9.5\,V\,+\,B0\,V   & 3       & 6     & 6   & X-ray pulsations \\
HD\,165052     & O7\,Vz\,+\,O7.5\,Vz & 300     & 1.2   & 7   & colliding wind binary \\
HD\,166033     & B1\,V               & 800     & 1.5   & 4,8 & \\
HD\,315032     & B2\,Vne             & 10      & 1.2   & 9   & \\
HD\,345439     & B1/B2\,V            & $<0.05$ & 1.5   & 10  & $\sigma$\,Ori\,E-analog \\
\hline
CPD\,-59$^{\circ}$\,2624 & O9.5\,V             & 1.9     & 2.3   & 11  & \\
HD\,37020      & B0.5\,V             & 25      & 0.45  & 12  & \\
HD\,46056      & O8\,V               & 32.5    & 1.4   & 13  & \\ 
HD\,46106      & O9.7\,III           & 0.9     & 1.4   & 13  & \\ 
HD\,46149      & O8.5\,V((f))        & 1.1     & 1.4   & 13  & \\
HD\,46150      & O5\,V((f))          & 22      & 1.4   & 13  & \\
HD\,46202      & O9\,V((f))          & 1.3     & 1.4   & 13  & \\ 
HD\,46223      & O4\,V((f))          & 24      & 1.4   & 13  & \\
HD\,46328      & B0.7\,IV            & 3       & 0.42  & 14  & X-ray pulsations \\      
HD\,60848      & O8:V:               & 0.13    & 0.5   & 15  & \\
HD\,93027      & O9.5\,IV            & 1.3     & 2.3   & 11  & \\
HD\,93521      & O9\,Vp              & 1.1     & 1.2   & 16  & \\
HD\,101008     & B1\,II/III          & 3       & 2.3   & 17  & \\ 
HD\,117357     & O9.5/B0\,V          & 20      & 7     & 18  & \\
HD\,125823     & B7\,IIIpv           & 0.005   & 0.14  & 19  & \\
HD\,152246     & O9\,IV              & 14      & 1.6   & 20  & triple \\
\hline
\end{tabular}
\tablefoot{
Distances and the interstellar absorption column density -- used to derive 
$L_{\rm X}$ -- are estimated on the basis of UBV photometry, or taken from 
the literature. \\
{\bf References:}
(1) De Becker et al.\ (\cite{DeBecker2005});
(2) Motch et al.\ (\cite{Motch2015});
(3) Torrej{\'o}n et al.\ (\cite{Torrejon2012});
(4) Oskinova et al.\ (in preparation);
(5) Fr\'emat et al.\ (\cite{Fremat2002});
(6) Trepl et al.\ (\cite{Trepl2012});
(7) Pittard \& Parkin (\cite{Pittard2010});
(8) Dahm et al.\ (\cite{Dahm2012});
(9) Damiani et al.\ (\cite{Damiani2004});
(10) Eikenberry et al.\ (\cite{Eikenberry2014});
(11) Naz\'e et al.\ (\cite{Naze2011});
(12) Stelzer et al.\ (\cite{Stelzer2005});
(13) Wang et al.\ (\cite{Wang2008});
(14) Oskinova et al.\ (\cite{Oskinova2014});
(15) Rauw et al.\ (\cite{Rauw2013});
(16) Rauw et al.\ (\cite{Rauw2012});
(17) Naz\'e et al.\ (\cite{Naze2013});
(18) Beer (\cite{Beer1961});
(19) Naz\'e et al.\ (\cite{Naze2014});
(20) Nasseri et al.\ (\cite{Nasseri2014}).
}
\end{table*}

Although strong and hard X-ray emission is often used as an indirect 
indicator of the presence of a magnetic field in massive stars
(Ignace et al.\ \cite{Ignace2013}; Naz\'e et al.\ \cite{Naze2014};
ud-Doula \& Naz\'e \cite{udDoulaNaze2015}),
previous studies have demonstrated that this condition 
is not a necessary indicator of stellar magnetism, and that magnetic 
stars can be insignificant X-ray sources
(Oskinova et al.\ \cite{Oskinova2011}; Naz\'e et al.\ \cite{Naze2014}).
Using the FORS\,2 observations, we intended to 
check whether strong X-ray emission always results from a magnetic field
and thus may be a good magnetic field indicator.

For this purpose, we extensively searched X-ray archives and selected 
early B-type stars with unusual X-ray characteristics: HD\,166033, 
HD\,110432, HD\,120991, and HD\,120324 (see Table\,\ref{tab:bstar}).
A detailed discussion 
of the X-ray properties of these stars will be presented in a forthcoming 
publication (Oskinova et al., in preparation).
We also obtained X-ray observations of 
the known magnetic B-type star HD\,345439 with the {\em XMM-Newton} telescope.
Below, we briefly discuss the results 
of our search for a correlation between X-rays and magnetic properties
for the objects described in this study.
 
HD\,166033 is thought to be the main ionizing source of the nebula around 
IC\,1274 (Dahm et al.\ \cite{Dahm2012}). The high X-ray luminosity of this 
B1V star is outstanding (however, one has to keep in mind the possible 
uncertainty on the distance). E.g.\ the X-ray luminosity of the well known 
magnetic stars $\tau$\,Sco (B0.2V) and $\xi^1$\,CMa (B0.7IV) are two orders of 
magnitude lower (Oskinova et al.\ \cite{Oskinova2011,Oskinova2014}). Yet, no 
magnetic field was detected in HD\,166033.

Among our targets there is also HD\,110432, which is a hard and bright X-ray source. 
This Be star is rotating near critical velocity, and is classified as a 
$\gamma$\,Cas-analog (e.g.\ Motch et al.\ \cite{Motch2015}, and references 
therein).
It was suggested that a magnetic field may be responsible for the 
unusual X-ray properties of these enigmatic objects (Smith et al.\ \cite{Smith2015}). 
However, our measurements do not confirm the presence of a strong magnetic field 
in this object.
Here, we report that its spectroscopic twin, HD\,120991, has 
similar X-ray properties, and  propose HD\,120991
as a $\gamma$\,Cas-analog.
Similarly, the X-ray luminosity 
of another of our targets, HD\,315032, is comparable to that of the $\gamma$\,Cas-analogs,
and its low signal-to-noise X-ray spectrum indicates that this star 
too is a hard X-ray source.
Further studies are needed to confirm it as a $\gamma$\,Cas-analog.
No strong magnetic field was detected in these objects either.   

The X-ray properties of the $\gamma$\,Cas-analogs are in sharp contrast with 
those of HD\,120324, another fast rotating Be star.
Its X-ray luminosity is low and the spectrum is quite soft.
Our study allows us to rule 
out a strong stellar magnetic field as a reason for this discrepancy. 

The fast rotating star HD\,345439 is an analog of 
$\sigma$\,Ori\,E, but its X-ray luminosity is about two orders of magnitude lower.
It is comparable to that of the magnetic B stars V1046\,Ori, $\zeta$\,Cas, V2052\,Ori, and LP\,Ori.
It was shown in the past that magnetic early-type stars can be 
insignificant X-ray sources (e.g., Oskinova et al.\ \cite{Oskinova2011}; 
Ignace et al.\ \cite{Ignace2013}).  
We should note that $\sigma$\,Ori\,E almost certainly has a low mass active companion
(e.g., Sanz-Forcada et al.\ \cite{SanzForcada2004}; Bouy et al.\ \cite{Bouy2009}).

BD\,$-$12\,4982 is a likely member of the NGC\,6604 cluster and one of its 
brightest X-ray sources.
To estimate its X-ray luminosity, we adopted the 
cluster distance and reddening from Kharchenko et al.\ (\cite{Kharchenko2009}).
It appears that this object has unremarkable X-ray properties. 
A magnetic field in BD\,$-$12\,4982 was measured in the whole spectrum,
at the 4.7$\sigma$ and 5.2$\sigma$ levels.

The X-ray emission from HD\,164816 was studied in detail by Trepl et al.\ 
(\cite{Trepl2012}), who detected X-ray pulsations with a period of 9.78\,s in 
this object.
The faint close companion to HD\,164816 found by Sana et al.\ (\cite{Sana2014}) 
could be an active pre-main sequence star responsible for the X-ray emission
(see also Sect.~\ref{sect:binarity}).
No magnetic field was detected in this source.

HD\,165052 is another massive binary. Presumably, the bulk of its X-ray 
emission is produced in the colliding winds of the two O stars 
(Pittard \& Parkin \cite{Pittard2010}).
Also in this source we did not detect a magnetic field.

An overview about the known X-ray properties of the 71 objects in our
survey is given in Table~\ref{tab:bstar}.
In Columns~1 and 2, we list the object name and the spectral type, followed
by the X-ray luminosity in Col.~3, and the adopted distance in Col.~4.
Col.~5 gives the corresponding reference and the last column comments on the object.
The upper part of the table lists objects studied in this article, the
lower part objects reported in Fossati et al.\ (\cite{Fossati2015}).
No details on X-rays are reported for the sources not listed in Table~\ref{tab:bstar}.
However, even this limited study confirms previous reports that while
some magnetic stars are strong X-ray sources, others are not.
Moreover, strong and hard X-rays 
from presumably single massive stars are not necessarily associated with the 
presence of stellar magnetic fields, and hence cannot serve as their indirect indicators
(e.g.\ Petit et al.\ \cite{Petit2013}).

\section{Discussion and conclusion}
\label{sect:discussion}

The ``B fields in OB stars'' collaboration aims at characterizing the incidence of magnetic fields 
in slowly rotating massive stars.
As part of this study, we employed the low-resolution FORS\,2 spectropolarimeter to observe a total
of 71 OB stars,
focussing on the fainter, less studied, objects and the faster rotators in our sample.
From the 32 observations of 28 OB stars reported in this article,
we were able to monitor the magnetic fields in CPD\,$-$57$^{\circ}$\,3509 (Przybilla et al.\ \cite{Przybilla2016})
and HD\,164492C (Hubrig et al.\ \cite{Hubrig2014a}),
confirm the magnetic field in HD\,54879 (Castro et al.\ \cite{Castro2015}),
and detect a magnetic field in CPD\,$-$62$^{\circ}$\,2124 (Castro et al., submitted).
While the standard analysis of HD\,345439 did not reveal a magnetic field,
the individual subexposures indicate that HD\,345439 may host a strong magnetic field that rapidly varies over 88\,min
(Hubrig et al.\ \cite{Hubrig2015a}).
A magnetic field in BD\,$-$12\,4982 was measured in the whole spectrum,
at 4.7$\sigma$ and 5.2$\sigma$ levels by the two groups,
while the measurements using the hydrogen lines showed only
significance levels below 3$\sigma$.

Since the BOB Collaboration decided to consider a magnetic field to be detected with FORS\,2
in a single observation only above the 5$\sigma$ level, and
excluding the two previously known magnetic stars HD\,46328 and HD\,125823,
we have detected with FORS\,2 magnetic fields in four stars in our sample of 69 OB stars:
CPD\,$-$57$^{\circ}$\,3509, HD\,164492C, HD\,54879, and CPD\,$-$62$^{\circ}$\,2124.
This leads to a magnetic field detection rate of $6\pm3\%$,
which is compatible with the detection rate of 7\% reported by Wade et al.\ (\cite{Wade2014}).
Please note however that CPD\,$-$57$^{\circ}$\,3509 and CPD\,$-$62$^{\circ}$\,2124 are both He-strong stars
and thus strongly magnetic (e.g.\ Smith \cite{Smith1996}).
Also the magnetic star in the system HD\,164492C is very likely He-strong
(Gonz\'alez et al., submitted).
For the 20 objects pre-selected on the basis of their low $v\,\sin\,i$,
we obtain a magnetic field detection rate of $5\pm5\%$.
An in-depth discussion on the incidence of magnetic fields in
all stars observed in our sample with both FORS\,2 and HARPSpol
will be presented in a forthcoming paper.

We again compared the results of the independent 
reduction and analysis carried out by two teams using different
and independent tools and pipelines and could show that the results agree
with expected statistical distributions.
This gives us high confidence of the accuracy of our longitudinal magnetic field measurements.

The results presented in this article underline the central role
of FORS\,2 observations for stellar magnetism studies
in stars of different spectral classification at almost all
stages of stellar evolution, especially for fainter targets.

\begin{acknowledgements}

LF acknowledges financial support from the Alexander von Humboldt Foundation.
AK acknowledges financial support from RFBR grant 16-02-00604A.
TM acknowledges financial support from Belspo for contract PRODEX {\sl Gaia}-DPAC.
MFN acknowledges support by the Austrian Science Fund (FWF)
in the form of a Meitner Fellowship under project number N-1868-NBL.
MS and SH thank Thomas Szeifert for providing the pipeline for the FORS spectra extraction.
This research has made use of the SIMBAD and ViZieR databases.

\end{acknowledgements}


\begin{thebibliography}{}

\bibitem[2015]{AertsRogers2015}
Aerts, C., \& Rogers, T.~M.\ 2015,
ApJL, 806, L33

\bibitem[1998]{Appenzeller1998}
Appenzeller, I., Fricke, K., F{\"u}rtig, W., et al.\ 1998,	
{The Messenger}, 94, 1

\bibitem[2002]{Bagnulo2002}
Bagnulo, S., Szeifert, T., Wade, G.~A., Landstreet, J.~D., \& Mathys, G.\ 2002,
A\&A, 389, 191

\bibitem[2009]{Bagnulo2009}
Bagnulo, S., Landolfi, M., Landstreet, J.~D., et al.\ 2009,
PASP, 121, 993

\bibitem[2012]{Bagnulo2012}
Bagnulo, S., Landstreet, J. D., Fossati, L., \& Kochukhov, O. 2012,
A\&A, 538, A129

\bibitem[2015]{Bagnulo2015}
Bagnulo, S., Fossati, L., Landstreet, J.~D., \& Izzo, C.\ 2015,
A\&A, 583, A115

\bibitem[1961]{Beer1961}
Beer, A.\ 1961,
MNRAS, 123, 191

\bibitem[2011]{Blomme2011}
Blomme, R., Mahy, L., Catala, C., et al.\ 2011,			
A\&A, 533, A4

\bibitem[2012]{BochkarevKaritskaya2012}
Bochkarev, N.~G., \& Karitskaya, E.~A.\ 2012,
in From Interacting Binaries to Exoplanets: Essential Modeling Tools,
eds.\ M.~T.~Richards, \& I.~Hubeny, IAU Symp., 282, 75

\bibitem[2009]{Bouy2009}
Bouy, H.,  Hu{\'e}lamo, N.,  Mart{\'{\i}}n, E.~L., et al.\ 2009,		
A\&A, 493, 931

\bibitem[2011]{Briquet2011}
Briquet, M., Aerts, C., Baglin, A., et al.\ 2011,		
A\&A, 527, A112

\bibitem[2011]{Brott2011}
Brott, I., de Mink, S.~E., Cantiello, M., et al.\ 2011,
A\&A, 530, A115

\bibitem[2015]{Buysschaert2015}
Buysschaert, B., Aerts, C., Bloemen, S., et al.\ 2015,		
MNRAS, 453, 89

\bibitem[2009]{Bychkov2009}
Bychkov, V.~D., Bychkova, L.~V., \& Madej, J.\ 2009,
MNRAS, 394, 1338

\bibitem[2002]{Carrier2002}
Carrier, F., North, P., Udry, S., \& Babel, J.\ 2002,
A\&A, 394, 151

\bibitem[2015]{Castro2015}
Castro, N., Fossati, L., Hubrig, S., et al.\ 2015,		
A\&A, 581, A81

\bibitem[2012]{Dahm2012}
{Dahm}, S.~E., {Herbig}, G.~H., \& {Bowler}, B.~P.\ 2012,
AJ, 143, 3

\bibitem[2004]{Damiani2004}
{Damiani}, F., {Flaccomio}, E., {Micela}, G., et al.\ 2004,		
ApJ, 608, 781

\bibitem[2005]{DeBecker2005}
De Becker, M., Rauw, G., Blomme, R., et al.\ 2005,			
A\&A, 437, 1029

\bibitem[2010]{Degroote2010}
Degroote, P., Briquet, M., Auvergne, M., et al.\ 2010,			
A\&A, 519, A38

\bibitem[2014]{Eikenberry2014}
Eikenberry, S.~S., Chojnowski, S.~D., Wisniewski, J., et al.\ 2014,		
ApJL, 784, L30

\bibitem[2009]{Ferrario2009}
Ferrario, L., Pringle, J.~E., Tout, C.~A., \& Wickramasinghe, D.~T.\ 2009,
MNRAS, 400, L71

\bibitem[2015]{Fossati2015}
Fossati, L., Castro, N., Sch\"oller M., et al.\ 2015,			
A\&A, 582, A45

\bibitem[2002]{Fremat2002}
Fr{\'e}mat, Y., Zorec, J., Hubert, A.-M., et al.\ 2002,			
A\&A, 385, 986

\bibitem[2013]{GrunhutWade2013}
Grunhut, J.~H., \& Wade, G.~A.\ 2013,
in Setting a New Standard in the Analysis of Binary Stars,
eds.\ K.~Pavlovski, A.~Tkachenko, \& G.~Torres, EAS Publ.\ Ser., 64, 67

\bibitem[2005]{Heger2005}
Heger, A., Woosley, S.~E., \& Spruit, H.~C.\ 2005,
ApJ, 626, 350

\bibitem[1994]{Houk1994}
Houk, N.\ 1994,
in The MK Process at 50 Years: A Powerful Tool for Astrophysical Insight,
eds.\ C.~J.~Corbally, R.~O.~Gray, \& R.~F.~Garrison,
Astr.\ Soc.\ of the Pac.\ Conf.\ Ser., 60, 285

\bibitem[1993]{HowarthReid1993}
Howarth, I.~D., \& Reid, A.~H.~N.\ 1993,
A\&A, 279, 148

\bibitem[1997]{Howarth1997}
Howarth, I.~D., Siebert, K.~W., Hussain, G.~A.~J., \& Prinja, R.~K.\ 1997,
MNRAS, 284, 265

\bibitem[2004a]{Hubrig2004a}
Hubrig, S., Kurtz, D.~W., Bagnulo, S., et al.\ 2004a,			
A\&A, 415, 661

\bibitem[2004b]{Hubrig2004b}
Hubrig, S., Szeifert, T., Sch\"oller, M., Mathys, G., \& Kurtz, D.~W.\ 2004b,
A\&A, 415, 685

\bibitem[2009]{Hubrig2009}
Hubrig, S., Sch{\"o}ller, M., Savanov, I., et al.\ 2009,		
Astr.\ Nachr., 330, 708

\bibitem[2011]{Hubrig2011b}
Hubrig, S., Ilyin, I., Briquet, M., et al.\ 2011,			
A\&A, 531, L20

\bibitem[2014]{Hubrig2014a}
{Hubrig}, S., {Fossati}, L., {Carroll}, T.~A., et al.\ 2014,		
A\&A, 564, {L10}

\bibitem[2015]{Hubrig2015a}
{Hubrig}, S., {Sch{\"o}ller}, M., {Fossati}, L., et al.\ 2015,		
A\&A, 578, {L3}

\bibitem[2016]{Hubrig2016}
Hubrig, S., Scholz, K., Hamann, W.-R., et al.\ 2016,			
MNRAS, 458, 3381

\bibitem[2013]{Ignace2013}
{Ignace}, R., {Oskinova}, L.~M., \& {Massa}, D.\ 2013,
MNRAS, 429, 516

\bibitem[2010]{Karitskaya2010}
Karitskaya, E.~A., Bochkarev, N.~G., Hubrig, S., et al.\ 2010,		
IBVS, 5950, 1

\bibitem[2009]{Kharchenko2009}
Kharchenko, N.~V., Piskunov, A.~E., R{\"o}ser, S., et al.\ 2009,	
A\&A, 504, 681

\bibitem[2002]{KoenEyer2002}
Koen, C., \& Eyer, L.\ 2002,
MNRAS, 331, 45

\bibitem[2011]{LeBouquin2011}
Le~Bouquin, J.-B., Berger, J.-P., Lazareff, B., et al.\ 2011,
A\&A, 535, A67

\bibitem[2003]{Lenzen2003}
Lenzen, R., Hartung, M., Brandner, W., et al.\ 2003,
SPIE Conf.\ Ser., 4841, 944

\bibitem[2005]{MaederMeynet2005}
Maeder, A., \& Meynet, G.\ 2005,
A\&A, 440, 1041

\bibitem[2016]{MandeldeMink2016}
Mandel, I., \& de Mink, S.~E.\ 2016,
MNRAS, 458, 2634

\bibitem[2016]{Marchant2016}
Marchant, P., Langer, N., Podsiadlowski, P., Tauris, T.~M., \& Moriya, T.~J.\ 2016,
A\&A, 588, A50

\bibitem[2001]{Mason2001}
Mason, B.~D., Wycoff, G.~L., Hartkopf, W.~I., et al.\ 2001,
AJ, 122, 3466

\bibitem[2006]{Mermilliod2006}
Mermilliod, J.~C.\ 2006,
VizieR Online Data Catalog: Homogeneous Means in the UBV System, 2168

\bibitem[2011]{Meynet2011}
Meynet, G., Eggenberger, P., \& Maeder, A.\ 2011,
A\&A, 525, L11

\bibitem[2008]{Morel2008}
Morel, T., Hubrig, S., \& Briquet, M.\ 2008,
A\&A, 481, 453

\bibitem[2014]{Morel2014}
Morel, T., Castro, N., Fossati, L., et al.\ 2014,			
{The Messenger}, 157, 27

\bibitem[2015]{Morel2015}
Morel, T., Castro, N., Fossati, L., et al.\ 2015,			
in {New Windows on Massive Stars}, eds.\ G.~Meynet, C.~Georgy, J.~Groh, \& P.~Stee,
IAU Symp., 307, 342

\bibitem[1978]{MorrisonConti1978}
Morrison, N.~D., \& Conti, P.~S.\ 1978,
ApJ, 224, 558

\bibitem[2015]{Motch2015}
Motch, C., Lopes de Oliveira, R., \& Smith, M.~A.\ 2015,
ApJ, 806, 177

\bibitem[2014]{Nasseri2014}
Nasseri, A., Chini, R., Harmanec, P., et al.\ 2014,			
A\&A, 568, A94

\bibitem[2011]{Naze2011}
Naz{\'e}, Y., Broos, P.~S., Oskinova, L., et al.\ 2011,			
ApJS, 194, 7

\bibitem[2013]{Naze2013}
Naz{\'e}, Y., Rauw, G., Sana, H., \& Corcoran, M.~F.\ 2013,
A\&A, 555, A83

\bibitem[2014]{Naze2014}
Naz{\'e}, Y., Petit, V., Rinbrand, M.~A, et al.\ 2014,		
ApJS, 215, 10

\bibitem[2012]{Neiner2012}
Neiner, C., Landstreet, J.~D., Alecian, E., et al.\ 2012,	
A\&A, 546, A44

\bibitem[2011]{Oskinova2011}
{Oskinova}, L.~M., {Todt}, H., {Ignace}, R., et al.\ 2011,		
MNRAS, 416, 1456

\bibitem[2014]{Oskinova2014}
{Oskinova}, L.~M., {Naz{\'e}}, Y., {Todt}, H., et al.\ 2014,		
Nature Comm., 5, 4024

\bibitem[2013]{Petit2013}
Petit, V., Owocki, S.~P., Wade, G.~A., et al.\ 2013,
MNRAS, 429, 398

\bibitem[2008]{PigulskiPojmanski2008}
Pigulski, A., \& Pojma{\'n}ski, G.\ 2008,
A\&A, 477, 917

\bibitem[2010]{Pittard2010}
{Pittard}, J.~M., \& {Parkin}, E.~R.\ 2010,
MNRAS, 403, 1657

\bibitem[2004]{Pourbaix2004}
Pourbaix, D., Tokovinin, A.~A., Batten, A.~H., et al.\ 2004,
A\&A, 424, 727

\bibitem[2016]{Przybilla2016}
Przybilla, N., Fossati, L., Hubrig, S., et al.\ 2016,			
A\&A, 587, A7

\bibitem[2012]{Rauw2012}
Rauw, G., Morel, T., \& Palate, M.\ 2012,
A\&A, 546, A77

\bibitem[2013]{Rauw2013}
Rauw, G., Naz{\'e}, Y., Spano, M., et al.\ 2013,			
A\&A, 555, L9

\bibitem[2013]{Rogers2013}
Rogers, T.~M., Lin, D.~N.~C., McElwaine, J.~N., \& Lau, H.~H.~B.\ 2013,
ApJ, 772, 21

\bibitem[2003]{Rousset2003}
Rousset, G., Lacombe, F., Puget, P., et al.\ 2003,
SPIE Conf.\ Ser., 4839, 140

\bibitem[2006]{Saesen2006}
Saesen, S., Briquet, M., \& Aerts, C.\ 2006,
Comm.\ in Asteroseismology, 147, 109

\bibitem[2014]{Sana2014}
Sana, H., Le Bouquin, J.-B., Lacour, S., et al.\ 2014,
ApJS, 215, 15

\bibitem[2004]{SanzForcada2004}
Sanz-Forcada, J., Franciosini, E., \& Pallavicini, R.\ 2004,
A\&A, 421, 715

\bibitem[2006]{Schnerr2006}
Schnerr, R.~S., Verdugo, E., Henrichs, H.~F., \& Neiner, C.\ 2006,
A\&A, 452, 969

\bibitem[2007]{Schoeller2007}
Sch{\"o}ller, M.\ 2007,
New Astronomy Reviews, 51, 628

\bibitem[2012]{Schoeller2012}
Sch{\"o}ller, M., Correia, S., Hubrig, S., \& Kurtz, D.~W.\ 2012,
A\&A, 545, A38

\bibitem[2014]{SimonDiazHerrero2014}
Sim{\'o}n-D{\'{\i}}az, S., \& Herrero, A.\ 2014,
A\&A, 562, A135

\bibitem[1996]{Smith1996}
Smith, K.~C.\ 1996,
Ap\&SS, 237, 77

\bibitem[2016]{Smith2015}
Smith, M.~A., Lopes de Oliveira, R., \& Motch, C.\ 2016,
Advances in Space Research, 58, 782

\bibitem[2008]{Snik2008}
Snik, F., Jeffers, S., Keller, C., et al.\ 2008,			
SPIE Conf.\ Ser., 7014, 70140O

\bibitem[2014]{Sota2014}
Sota, A., Ma{\'{\i}}z Apell{\'a}niz, J., Morrell, N.~I., et al.\ 2014,	
ApJS, 211, 10

\bibitem[2005]{Stelzer2005}
Stelzer, B., Flaccomio, E., Montmerle, T., et al.\ 2005,			
ApJS, 160, 557

\bibitem[1997]{Stickland1997}
Stickland, D.~J., Lloyd, C., \& Koch, R.~H.\ 1997,
{The Observatory}, 117, 295

\bibitem[2006]{Telting2006}
Telting, J.~H., Schrijvers, C., Ilyin, I.~V., et al.\ 2006,		
A\&A, 452, 945

\bibitem[2012]{Torrejon2012}
Torrej{\'o}n, J.~M., Schulz, N.~S., \& Nowak, M.~A.\ 2012,
ApJ, 750, 75

\bibitem[2012]{Trepl2012}
{Trepl}, L., {Hambaryan}, V.~V., {Pribulla}, T., et al.\ 2012,		
MNRAS, 427, 1014

\bibitem[2010]{Tuthill2010}
Tuthill, P., Lacour, S., Amico, P., et al.\ 2010,
SPIE Conf.\ Ser., 7735, 77351O

\bibitem[2008]{udDoula2008}
ud-Doula, A., Owocki, S.~P., \& Townsend, R.~H.~D.\ 2008,
MNRAS, 385, 97

\bibitem[2016]{udDoulaNaze2015}
ud-Doula, A., \& Naze, Y.\ 2016,
Advances in Space Research, 58, 680

\bibitem[2012]{Wade2012}
Wade, G.~A., Grunhut, J., Gr{\"a}fener, G., et al.\ 2012,		
MNRAS, 419, 2459

\bibitem[2014]{Wade2014}
Wade, G.~A., Petit, V., Grunhut, J., \& Neiner, C.\ 2014,
arXiv:1411.6165

\bibitem[2016]{Wade2016}
Wade, G.~A., Neiner, C., Alecian, E., et al.\ 2016,
MNRAS, 456, 2

\bibitem[2008]{Wang2008}
Wang, J., Townsley, L.~K., Feigelson, E.~D., et al.\ 2008,			
ApJ, 675, 464

\bibitem[2015]{Wisniewski2015}
Wisniewski, J.~P., Chojnowski, S.~D., Davenport, J.~R.~A., et al.\ 2015,	
ApJL, 811, L26

\bibitem[2006]{Yoon2006}
Yoon, S.-C., Langer, N., \& Norman, C.\ 2006,
A\&A, 460, 199

\end{thebibliography}
\end{document}